\documentclass[sigconf, screen]{acmart}

\AtBeginDocument{%
  }

\copyrightyear{2025}
\acmYear{2025}
\setcopyright{cc}
\setcctype[4.0]{by}
\acmConference[CHI '25]{CHI Conference on Human Factors in Computing Systems}{April 26-May 1, 2025}{Yokohama, Japan}
\acmBooktitle{CHI Conference on Human Factors in Computing Systems (CHI '25), April 26-May 1, 2025, Yokohama, Japan}\acmDOI{10.1145/3706598.3713180}
\acmISBN{979-8-4007-1394-1/25/04}

\usepackage{graphicx}
\usepackage[noabbrev,capitalize]{cleveref}
\usepackage{xspace}
\usepackage{enumitem}
\begin{document}
\newcommand{\system}{PlantPal\xspace}
\renewcommand{\sectionautorefname}{Section}
\renewcommand{\subsectionautorefname}{Section}
\renewcommand{\subsubsectionautorefname}{Section}
\renewcommand*{\appendixautorefname}{Appendix}
\renewcommand*{\figureautorefname}{Fig.}
\title{\system: Leveraging Precision Agriculture Robots to Facilitate Remote Engagement in Urban Gardening}
\author{Albin Zeqiri}
\email{albin.zeqiri@uni-ulm.de}
\orcid{0000-0001-6516-3810}
\affiliation{%
  \institution{Institute of Media Informatics, Ulm University}
  \city{Ulm}
  \country{Germany}
}

\author{Julian Britten}
\email{julian.britten@uni-ulm.de}
\orcid{0000-0002-2646-2727}
\affiliation{%
  \institution{Institute of Media Informatics, Ulm University}
  \city{Ulm}
  \country{Germany}
}

\author{Clara Schramm}
\email{clara.schramm@uni-ulm.de}
\orcid{0009-0009-2297-5304}
\affiliation{%
  \institution{Institute of Media Informatics, Ulm University}
  \city{Ulm}
  \country{Germany}
}

\author{Pascal Jansen}
\email{pascal.jansen@uni-ulm.de}
\orcid{0000-0002-9335-5462}
\affiliation{%
  \institution{Institute of Media Informatics, Ulm University}
  \city{Ulm}
  \country{Germany}
}

\author{Michael Rietzler}
\email{michael.rietzler@uni-ulm.de}
\orcid{0000-0003-2599-8308}
\affiliation{%
  \institution{Institute of Media Informatics, Ulm University}
  \city{Ulm}
  \country{Germany}
}

\author{Enrico Rukzio}
\email{enrico.rukzio@uni-ulm.de}
\orcid{0000-0002-4213-2226}
\affiliation{%
  \institution{Institute of Media Informatics, Ulm University}
  \city{Ulm}
  \country{Germany}
}

\renewcommand{\shortauthors}{Zeqiri et al.}
\renewcommand{\shorttitle}{PlantPal: Precision Agriculture Robots for Urban Gardening}

\begin{abstract}
Urban gardening is widely recognized for its numerous health and environmental benefits. However, the lack of suitable garden spaces, demanding daily schedules and limited gardening expertise present major roadblocks for citizens looking to engage in urban gardening. While prior research has explored smart home solutions to support urban gardeners, these approaches currently do not fully address these practical barriers. In this paper, we present PlantPal, a system that enables the cultivation of garden spaces irrespective of one's location, expertise level, or time constraints. PlantPal enables the shared operation of a precision agriculture robot (PAR) that is equipped with garden tools and a multi-camera system. Insights from a 3-week deployment (N=18) indicate that PlantPal facilitated the integration of gardening tasks into daily routines, fostered a sense of connection with one's field, and provided an engaging experience despite the remote setting. We contribute design considerations for future robot-assisted urban gardening concepts.
\end{abstract}



\begin{teaserfigure}
  \includegraphics[width=\textwidth]{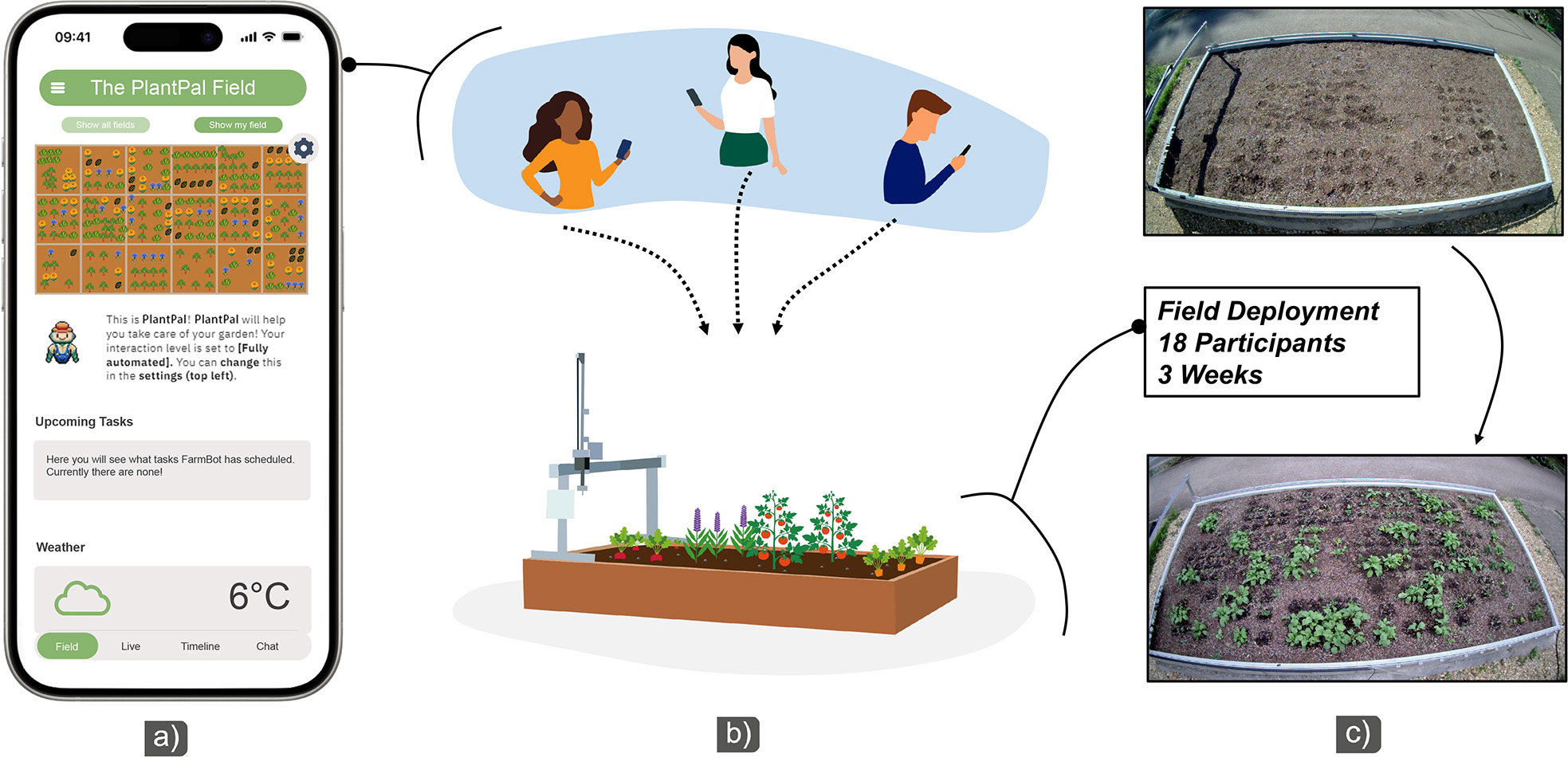}
  \caption{\system is a web application that allows for shared control of a precision agriculture robot (PAR) to enable remote robot-assisted urban gardening for multiple users. a) \system features various field views, dynamic field visualizations, live streams, timelines, and a chat. b) Each user of \system is assigned to their own field plot (a 1m x 1m space) and can remotely execute tasks (e.g., sowing seeds, watering, and weeding) by sending requests to FarmBot~\cite{FarmBotMain}, an open-source PAR, that was installed on a real garden bed (18m²). c) We deployed \system in a 3-week evaluation. After the study period, most participants successfully cultivated various crops on their plots.}
  \Description{This is the teaser figure showing the main views of the interface implemented for remote robot-assisted urban gardening. On the right, one can see the results from our study, which were that multiple participants were able to cultivate plants on a garden bed over three weeks using our system}
  \label{fig:teaser}
\end{teaserfigure}


\maketitle

\section{Introduction}\label{sec:introduction}
Engagement in urban gardening is considered an effective method to preserve local biodiversity~\cite{clucas2018,tresch2019,goddard2010}, enhance access to fresh produce~\cite{opitz2016,garcia2018}, and improve citizens' mental and physical well-being~\cite{soga2017,wakefield2007, brown2000}. Many large-scale political initiatives (e.g.,~\cite{EU, USEPA}) highlight urban gardening as a key component of broader efforts to expand and enhance green spaces in densely populated cities. Consequently, various research fields, including Human-Computer Interaction (HCI)~\cite{BiggsUG2021,agricultureImaginarySoden,Rodgers2020,community,smartWatering2}, have investigated strategies to promote urban gardening and re-engage former gardeners.\\

Prior HCI research surrounding the garden has focused on assisting individuals in developing gardening skills (e.g.,~\cite{IDROPO,growkit,wesenseEdu}), exploring hidden or unnoticed aspects of their gardens (e.g.,~\cite{AmbientBirdHouse,CameraTrap,BirdHouseCompanion}), and fostering social connections among gardeners (e.g.,~\cite{smartWatering2,urbanGardeningGroupsCompanion}). Numerous studies have proposed smart automation or sensor kits (e.g.,~\cite{smartWatering1,smartWatering2,urbanGardeningGroupsCompanion,growkit}) to enhance precision in plant care and improve gardeners' expertise by providing additional information~\cite{Rodgers2020}. Approaches leveraging smart gardening devices have likewise been proposed to facilitate collaboration and task management in community gardens~\cite{agricultureImaginarySoden,Rodgers2020,CommunityGardenParticipatory}. 
While these strategies effectively support individuals who are already regularly involved in gardening, they fall short of making consistent engagement in gardening more approachable or feasible for the broader population. As related research suggests, primary barriers for citizens interested in gardening extend beyond lacking knowledge or coordination~\cite{schupp2016, goodfellowbarriers2022}. Especially in confined urban environments, they include practical challenges such as lacking availability of spaces suitable for urban gardening (i.e., backyards or balconies)~\cite{goodfellowbarriers2022, schupp2016, schuppsharp2012, conwaybarriers2016c, space_apparicio2013predictors, space_conway2014tending}, hectic daily schedules~\cite{conwaybarriers2016c, schupp2016}, limited tolerance for the physical demands of gardening~\cite{goodfellowbarriers2022}, or inconsistent motivation~\cite{conwaybarriers2016c, lee2019influence}. These factors, frequently in combination, have been shown to discourage individuals interested in gardening or those who have previously tried it from re-engaging~\cite{goodfellowbarriers2022}. The question of how technological advances can be leveraged to shape alternative urban gardening experiences that are more accessible to a broader audience remains under-addressed in current research.\\

In this paper, we introduce \system, a system designed to facilitate on-demand access to and the cultivation of garden spaces regardless of an individual's location, expertise, or time limitations. At the core of \system is a remotely controllable Precision Agriculture Robot~(PAR) named FarmBot~\cite{FarmBotMain}, equipped with essential gardening tools and resources, enabling it to perform a range of gardening tasks on demand (\autoref{fig:teaser}b). \system also features a multi-camera setup that provides real-time visual feedback, allowing users to verify the PAR's actions and monitor plant health and growth progress on the field. As users cultivate their gardens using \system, the sampled data are used to create digitally augmented visualizations, including detailed field views, plant growth time-lapses, and event timelines for one's field plot (\autoref{fig:teaser}a). Additionally, by utilizing a PAR as a shared resource among multiple users, \system allows up to 18 people to cultivate their garden plots simultaneously.
The design of \system followed a three-step process. We reviewed existent literature surrounding urban gardening, technology-mediated nature engagement in HCI, and Human-Robot Interaction (HRI). Based on this knowledge, we ideated by mapping PARs' capabilities against current challenges preventing participation in urban gardening. We conducted a formative survey (N=42) to probe stances toward aspects resulting from our initial mapping (e.g., remote engagement in urban gardening and collaboration with PARs). Leveraging the acquired feedback, we derived three design goals that guided the design and implementation of \system.\\

We deployed our prototype (\autoref{fig:teaser}c) on a real garden bed (18m²) during a 3-week field study (N=18) to understand how users engage with \system and probe how the introduction of remote interaction with a gardening bed using PARs affects users' connectedness to their plots, longitudinal engagement, gardening success, and perceptions urban gardening. 
Our findings indicate that \system facilitates the integration of garden cultivation into daily routines, provides an engaging experience, and increases gardeners' perceived connectedness to their fields, despite the remote setting. Additionally, we found trends suggesting that the degree to which PAR automation capabilities are leveraged may impact gardeners' perceived connectedness and longitudinal engagement with remote urban gardening. Based on the development and evaluation of \system, we derive design considerations relevant to the design of future PAR-enabled urban gardening concepts. In summary, we contribute the following:
\begingroup
\setlist[enumerate]{itemsep=0pt, topsep=2pt} 
\begin{enumerate}
    \item The development of \system, a proof-of-concept system leveraging shared control over a PAR to enable remote cultivation of a real garden plot for multiple users. The setup offers a flexible control approach between the user and the PAR, offering dynamic adaptability to individual schedules while digitally augmenting the visualization of plant growth to enhance engagement and accessibility.
    \item Insights from a 3-week exploratory deployment of \system (N=18) indicating gardening success, an engaging and satisfying user experience, and connectedness to a garden despite a fully remote setting.
    \item Design considerations for future PAR-enabled urban gardening concepts and fully remote technology-supported nature experiences, addressing free exploration, risk of destruction, sustainable resource use, and digital augmentation.
\end{enumerate}
\endgroup

\section{Background And Related Work}\label{sec:rel}
The development of \system is grounded in related research on challenges of citizens looking to pursue urban gardening, HCI approaches aiming to enhance gardening practices, and HRI with PARs. 
\subsection{Challenges Related To Urban Gardening}\label{subsec:rel-challenges} Urban gardening encompasses all the practices related to growing food within and near cities, from inner city allotments and community gardens to periurban off-ground cultivation~\cite{ernwein2014a}. Going by this definition, practices such as backyard, allotment, rooftop, balcony, and community gardening are included under the broader umbrella term of urban gardening. Engagement in urban gardening is connected to numerous benefits, such as enhancing well-being and food resilience~\cite{opitz2016,brown2000,garcia2018}. Much research on urban gardening focuses on understanding the motivations, strategies, goals, and challenges of citizens who actively engage in urban gardening or intend to do so (e.g.,~\cite{lewis2018, schupp2016,urbanGardeningGroupsCompanion,CommunityGardenParticipatory}). Prior work has identified a wide range of motivations, including practical, intrinsic, and aesthetic factors~\cite{murtagh2023}. According to ~\citet{murtagh2023}, practical motivations often center around food production and promoting biodiversity, while intrinsic motivations typically involve personal pleasure and enjoyment throughout the growing process. Additionally, aesthetic motivations encompass the desire to shape one’s environment and are known as key driving factors. Previous research has demonstrated that, despite strong motivations, the ability to act on intentions to cultivate an urban garden is often accompanied by various challenges~\cite{guitart2012, goddard2013barriersa, goodfellowbarriers2022}. A common issue is the lack of accessible spaces for urban gardening~\cite{conwaybarriers2016c, goddard2010, schupp2016, goodfellowbarriers2022, space_apparicio2013predictors, space_conway2014tending}. Previous studies have additionally noted the unequal distribution of green spaces between lower- and higher-income neighborhoods in large cities~\cite{unequal1,unequal2,unequal3}. Gardening also demands knowledge of crop seasonality, the required frequency of plant care tasks, and the ability to assess plant health throughout the growth cycle. Lacking such knowledge has been shown to impede crop cultivation success~\cite{goodfellowbarriers2022,cerda2022}, which can, in turn, diminish motivation, especially for novice gardeners~\cite{conwaybarriers2016c}. For those aiming to engage in urban gardening consistently, integrating this practice into their daily routines is a key consideration. Grassroots initiatives like urban community gardens, where individuals share gardening spaces, aim to reduce barriers and foster social connection~\cite{guitart2012, urbanGardeningGroupsCompanion}. Research indicates that interest in these gardens has increased recently despite a temporary decline during the COVID-19 pandemic~\cite{bieri2024a}. Community gardens provide opportunities for members to share knowledge, support those new to gardening, and manage tasks collaboratively, addressing challenges such as the lack of private green space and limited gardening experience. However, community gardens are always accessible~\cite{guitart2012}.

\subsection{Supporting Gardening Through Technology}\label{subsec:tech-garden-rel}
In HCI research, various works have focused on understanding and supporting gardeners, not just in urban settings. Research on urban gardening often includes ethnographic studies of practices, traditions, and challenges in private and community gardens (e.g.,~\cite{community,CommunityGardenParticipatory}). A key focus of this research has been exploring how technology can be introduced to better support gardeners in their activities~\cite{Rodgers2020}. Understanding where and when technology may enhance gardening or any nature engagement experience is crucial as the introduction of technology also has the potential to diminish nature experiences~\cite{jones2018dealing, cumboAdverse2020,bidwell2010}.
Since gardening encompasses activities and experiences that go beyond the cultivation of plants, technological support should extend to various aspects beyond the gardening process itself. 
In a literature mapping,\citet{Rodgers2020} show that technical approaches surrounding urban gardening primarily aim to teach gardening skills, support the connection and coordination between gardeners, and reduce resource waste. Technologies used to support urban gardeners and gardening communities frequently fall under the broader category of IoT technologies~\cite{Rodgers2020}, including smart irrigation systems~\cite{pearceSmartWatering, smartWatering1, smartWatering2} and sensory toolkits~\cite{growkit, wesenseEdu} to monitor plant health markers. For instance, GrowKit~\cite{growkit} or WeSense~\cite{wesenseEdu} leverage smart sensors to educate users on plant health.  With "Connected Roots", \citet{smartWatering2} demonstrated how automated irrigation systems linked across multiple units in a residential building can facilitate interactions among residents interested in gardening. This approach exemplifies how automation can be used to assign new social value to a typically repetitive gardening task. When and where to incorporate technology to support gardening experiences has also been a research concern in the past. 
Additionally, a growing body of research explores ways to strengthen human-nature relationships~\cite{Rodgers2020,Webber2023CHI}. For example, \citet{CameraTrap} demonstrated using camera traps to help citizens observe and reflect on backyard ecosystems, while \citet{AmbientBirdHouse} proposed using technology-mediated auditory experiences to raise awareness of local bird species and foster a sense of connectedness with local biodiversity~\cite{BirdHouseCompanion, AmbientBirdHouse}. These works do not directly address the topic of urban gardening in the sense of crop cultivation. However, they focus on non-human actors and habitats users may create through gardening.\\ 

In summary, prior work investigating how the introduction of technology may support the gardening endeavors of urban dwellers has focused on enhancing gardening skills, facilitating social interactions and coordination between gardeners, promoting sustainability through resource-efficient practices and monitoring tools, and aspects that go beyond the process of plant cultivation. In the context of urban gardening, technology has been successfully employed to enhance gardeners' capabilities. However, as highlighted in the previous section, many users express interest in gardening but are hindered by practical barriers. Addressing the needs of these individuals requires shifting from \textit{enhancing capabilities} to \textit{creating opportunities} through technology. 

\subsection{Collaboration with Agriculture Robots}\label{subsec:rel-robots}
Previous HCI research on enhancing gardening capabilities has primarily used traditional smart gardening devices. However, recent advances in precision agriculture enable technology to take a more active role in gardening. PARs are autonomous or semi-autonomous systems designed to perform tasks such as planting, watering, weeding, and monitoring crop health~\cite{bechar2021, vasconez2019}. The scale at which PARs are deployed, whether in small gardens or large fields, affects their size, functionality, and user interaction. Larger PARs manage extensive farm operations~\cite{PARLarge}, while smaller ones are suited for local urban settings~\cite{goddard2021}. Small-scale consumer PARs such as FarmBot are often fixed in private backyards and do not require significant movement outside the designated field. Remote controllability allows users to interact with these PARs on-demand, whether nearby or at a distance~\cite{vasconez2019}. Control methods for PARs vary depending on task complexity and environmental conditions~\cite{controlmethod1, controlmethod2tax}. Fully autonomous PARs handle simple tasks like irrigation, while more complex tasks use semi-autonomous control, where human intervention is needed for decision-making~\cite{vasconez2019}. High-precision tasks, such as pruning or inspection, rely on teleoperation, with human operators remotely guiding the PAR step by step. Similar mixed-initiative approaches are commonly found in HRI research~\cite{control3, controlmethod2tax}.
As previously mentioned, crop cultivation requires consistent management of gardening tasks. PARs are able to execute tasks like watering~\cite{irrigationSpraying,irrigationSpraying2}, pruning~\cite{pruning}, harvesting~\cite{bac2014harvesting,harvesting2}, monitoring~\cite{monitoring,lunadei2012monitoring}, and mapping~\cite{mapping}, often necessitating specialized hardware~\cite{vasconez2019}. Small-scale PARs, like those designed for individual consumers, are often built to handle various gardening tasks, prioritizing user convenience. The aforementioned factors (i.e., scale, interaction proximity, tooling) additionally influence how PARs visualize information for the user. Effective information communication is important for maintaining situation awareness, trust, and acceptance~\cite{hriMetrics} across diverse tasks, settings, and interaction strategies. Meta-analyses from HRI and HCI highlight the importance of minimalism and simplicity, ensuring consistency while delivering only relevant information~\cite{controlmethod1, controlmethod2tax}. The level of detail is largely influenced by the task, control strategy, and user expertise. For instance, users of commercial PARs may require less detailed information than remote operators managing large-scale farming tasks with drones.\\ 

In urban gardening, using PARs for collaborative interaction introduces novel concepts, such as fully remote engagement, due to the broad range of tasks PARs can manage. Prior work, such as \citet{Webber2023CHI}, comprehensively reviewed the literature on technology-mediated nature engagement, finding that approaches vary across the dimensions of distance and directness. In distant settings, engagement often involves interactive videos~\cite{mediatedNature}, abstract representations~\cite{abstractNature}, or computer-generated depictions~\cite{generatedNature}. Shared PARs represent a novel form of distant nature engagement, where remote engagement with a robotic actor leads to tangible physical changes in the environment. Further, discourse about the effects of PARs deployed at scale in future cities is already emerging (cf.~\cite{goddard2021}). Therefore, investigating interaction with PARs for urban gardening could open new research spaces and facilitate the design of novel urban greening strategies. The following sections detail how \system adopts this approach and addresses common barriers to urban gardening participation.

\section{Designing and Implementing \system}\label{sec:method-all}
In the following, we detail the user-centered design process~\cite{ucd} used to develop \system. We first describe the derivation of our initial concept and then explain the hardware setup and software implementation of \system.
\subsection{Design Rationale}\label{subsec:design-rationale}
By reviewing prior studies surrounding urban gardening, we provided an overview of barriers that prevent engagement with urban gardening. The development of \system aimed to create a novel PAR-enabled urban gardening experience for individuals interested in gardening but who find it too inaccessible or impractical to pursue. Shared control over a PAR to cultivate gardens remotely has remained unexplored in prior work (see \autoref{subsec:tech-garden-rel}), leaving the design of such a system ambiguous. 
We initially set out to define design goals to guide the development of \system. In this process, we began by mapping the capabilities of current PARs (see \autoref{subsec:rel-robots}) to the challenges urban gardeners face (see \autoref{subsec:rel-challenges}). The aim was to systematically align and ideate the technological possibilities of PARs with urban gardeners' real-world needs and challenges. Each mapping included an explanation of how a PAR capability could address a specific challenge (\textit{Challenge: Availability of suitable spaces $\rightarrow$ PAR capability: remote controllability $\rightarrow$ Leverage remote control capabilities to facilitate on-demand access to a field managed by a PAR}). Two authors first independently generated and iteratively refined these mappings, resolving conflicts through discussion.
Aside from mappings surrounding immediate and continuous interaction with a potential garden space, ambiguities remained around balancing automation, user control, and information communication. For example, remote access to a distant green space may bridge the unavailability of nearby green spaces. Still, preferences for active involvement may vary depending on users' gardening expertise, goals, motivations, and how comfortable they feel about having control over a shared PAR. 

\subsection{Formative Survey}\label{subsec:formative-survey}
We conducted an online formative survey (N=42) to gather user perspectives on remote collaboration with a PAR and preferences for addressing ambiguities in the previous challenge-capability mappings.

\subsubsection{Survey Design}\label{subsubsec:design-of-formative-survey}
After filling out consent forms, the formative survey began with demographic questions. Participants were then presented with a list of barriers identified from our review of urban gardening literature and asked to indicate which barrier primarily prevents them from engaging in urban gardening. The main section of the survey focused on understanding participants' views on collaborating with PARs to cultivate garden spaces remotely. Given that consumer PARs are not widely known, we included a segment with visual depictions of FarmBot, and its capabilities. We then introduced the concept of using PARs to enable remote experiences in a shared garden. This was followed by various statements gauging expectations regarding the role of PARs in decision-making, additional functionalities other than gardening tasks, and the potential for sharing the robot as a resource among multiple gardeners. Statements were rated on a 7-point Likert scale. Afterward, participants were asked to briefly describe concerns and opportunities they saw with our concept.

\subsubsection{Participants}\label{subsubsec:participants-of-formative-survey}
We recruited participants from Prolific\footnote{https://www.prolific.com/ Accessed: 24/01/25}, which has been shown to provide reliable data~\cite{peer2022} and further allows for participant filtering. Initially, 50 participants were recruited, of whom eight failed attention checks (according to Prolific guidelines\footnote{https://researcher-help.prolific.com/en/article/fb63bb Accessed: 24/01/2025}), leaving 42 participants. 
Their ages ranged from 20 to 71 years ($M=30.80$, $SD=10.06$). 22 participants identified as female, and 20 participants identified as male. Regarding occupation, 24 participants were employed, 7 were self-employed, 6 were students, 4 were out of work, and 1 was retired. Thirteen participants held a bachelor's degree, 12 completed high school, 12 held a master’s degree, and 5 held no formal degree. Participants resided in various living environments: 25 in urban areas, 13 in suburban areas, and 4 in rural areas. 
Our sample consisted of participants who \textit{(1) self-reported interest in pursuing gardening and (2) were unable to do so to their liking at the time of the study}. The most frequently reported barrier in our sample was lacking access to green spaces ($67.60\%$), similar to related literature (e.g.,~\cite{conwaybarriers2016c, goodfellowbarriers2022, goddard2013barriersa}).  

\subsubsection{Collaborating with PARs}\label{subsubsec:dg-stats}
We analyzed how user ratings were distributed to understand how preferences varied across previously identified context-dependent aspects of the proposed concept. Results indicate that preferences for control initiatives differed considerably. Statements proposing the PAR take full initiative in gardening tasks, with users as spectators, were more frequently met with reservations. More precisely, \textit{40.54\%} of participants agreed with this notion, while \textit{59.46\%} disagreed. In contrast, \textit{45.95\%} of participants preferred taking the initiative in gardening decisions, with the robot merely serving as an \textit{"extended arm"} to access gardening spaces, while \textit{29.73\%} disagreed and \textit{24.32\%} were undecided. Statements proposing a hybrid approach, where the robot simplifies repetitive tasks such as watering and reviews user actions, while the user takes the initiative for more complex tasks such as deciding on the removal of weeds, received the most agreement (\textit{75.68\%}). \textit{80.95\%} of participants welcomed the notion of using a PAR as a shared resource among multiple gardeners, while \textit{19.05\%} disagreed.\\

\subsubsection{Remote Garden Cultivation}\label{subsubsec:dg-qualitative}
Participants provided brief free-text responses about opportunities and concerns regarding remote garden cultivation. 
Our goal in analyzing the qualitative data was to identify expressed opportunities and concerns and assess their relevance. We did so by inductively coding responses as one or more keywords to summarize the main opportunities or concerns. Similar to~\citet{elliott2018a}, we then counted occurrences of keywords to indicate their relevance within our sample.
The coding was done in a joint session by two authors (result: 30 codes). Codes were discussed and merged in the same joint session, resulting in 17 codes. To convey the trends of our data, we present the most frequent sentiments supported by excerpts from the feedback and their incidence (see supplementary material for the full code list). Participants expected PAR-supported remote garden cultivation to potentially raise \textit{efficiency (30x)}, produce larger \textit{yields (10x)}, and increase \textit{accessibility (10x)}. While efficiency and accessibility of remotely managing a garden were appreciated ("\textit{[..] allows for more people to take control of growing their own plants/food etc. within society as there is a bit less maintenance and time required}" (P14)), reservations were also expressed (\textit{"It may induce dependence on technology, a lot of expenses, lack of direct benefits from gardening, perception of the nature as something totally controllable"} (P36)). \textit{Destruction (15x)} (\textit{"That it may malfunction and damage the beds and vegetables."} (P15)) and disconnectedness from nature (15x) (\textit{"Disconnection from nature due to less interaction with plants; expensive technology." (P38)}) were the most mentioned concerns. Notably, this has been highlighted as a concern in prior literature as well~\cite{jones2018dealing, Rodgers2020}. Participants further note that PAR-supported gardening would not be considered a replacement for gardening but rather an alternative (\textit{"I don't have a garden so I think it's useful for that but this is more like its own thing to me. I can garden but it's a different gardening"} (P41)). This alternative way of garden cultivation may change \textit{perceptions (10x)} of gardening to be \textit{"only about the result at the end." (P2)}.

\begin{figure*}[t]
    \centering
    \includegraphics[width=1\linewidth]{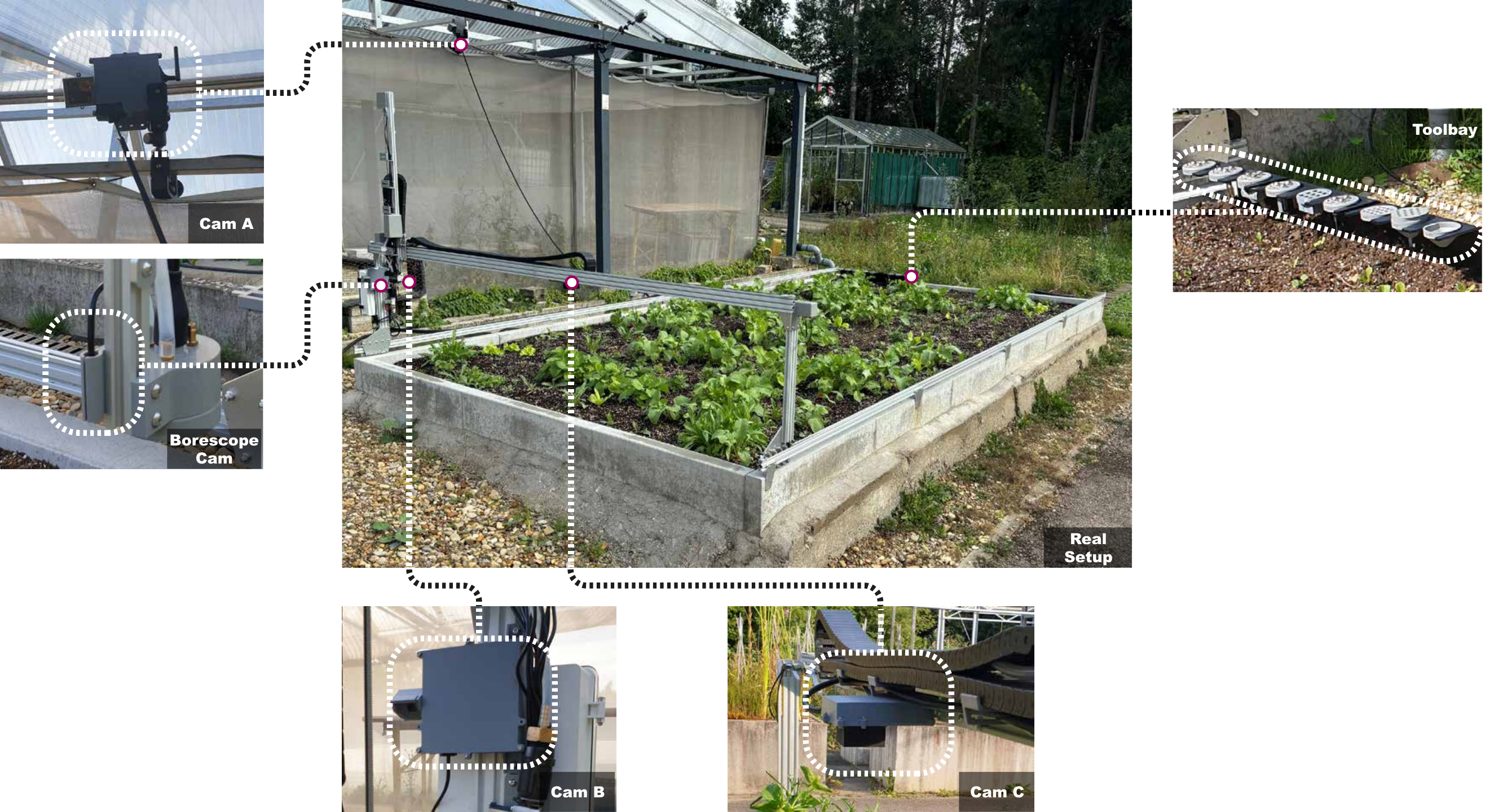}
    \caption{An illustration showing the hardware components of \system. We installed a PAR (FarmBot) on a real garden bed and extended its camera system (originally including only the borescope camera) to provide multi-view monitoring (cameras A, B, and C). FarmBot executes gardening tasks using various tools held in a tool bay at a fixed location on the garden bed.}
    \Description{Figure 2 shows the hardware setup for PlantPal. It shows that we installed a PAR called FarmBot on a real garden bed and extended its camera system to provide multi-view monitoring.}
    \label{fig:cams}
\end{figure*}

\subsubsection{Design Goals}\label{subsubsec:design-goals}

Our initial analysis of matching PAR capabilities and urban gardening challenges, along with preferences and concerns shared by users in the formative survey, informed the following design goals.

\subsubsection*{\textbf{D1 Remote On-Demand Access}} Spaces suitable for gardening are often not equally distributed~\cite{unequal1,unequal2,unequal3} or otherwise not accessible. From our initial matching of PAR capabilities with gardeners' challenges (cf. \autoref{subsec:design-rationale}), we conclude that \system should function as a fully remote concept on a consistently available private device. This concept can facilitate shared interaction with a distant garden irrespective of physical presence.

\subsubsection*{\textbf{D2 Adaptable Initiative in Decision-Making}} Preferences regarding automation and control initiative between a human user and PAR did not indicate that one specific control initiative was preferred over the other (cf. \autoref{subsubsec:dg-stats}). We conclude that \system should foster flexibility, allowing users to calibrate their preferred control initiative according to their goals, preferences, expertise, and contextual factors.

\subsubsection*{\textbf{D3 Meaning Beyond Execution of Gardening Tasks}} Participants voiced concerns about absent direct interaction with a garden leading to disengagement from nature and potential destruction (cf. \autoref{subsubsec:dg-qualitative}). They further view PAR-supported remote gardening as an alternative to traditional gardening rather than a replacement. We conclude that \system should provide ways to engage with gardening beyond the execution of plant care tasks. \system should aim to balance the introduced remoteness (\textit{D1}) and foster a sense of ownership and connection to one's garden space, avoiding adverse disengagement as noted in our formative survey and prior literature (e.g.,~\cite{bidwell2010, demaConcerns2019, cumboAdverse2020}).\\

The following sections outline the technical setup of \system and detail the software implementation, highlighting the integrated strategies and their alignment with the design goals.
\subsection{Technical Setup}\label{subsec:technical-setup}
As discussed in \autoref{subsec:rel-robots}, PARs designed for individual consumers are beginning to emerge. While many are still in the prototypical phase, their availability for open-source development makes them suitable platforms for \system. To enable remote on-demand garden interaction, PAR prototypes should support diverse gardening tasks and allow hardware and software extensions. We selected FarmBot (Genesis XL model), an open-source PAR, as the foundation for \system. FarmBot is designed for small-scale gardening and can execute seeding, watering, weeding, and sensing (e.g., soil moisture) actions~\cite{BotGeneral}. Its open-source framework allows for customization and integration of additional features, aligning with our purposes.

\subsubsection{Movement}\label{subsubsec:movement}
\begin{figure}[ht]
    \centering
    \includegraphics[width=1\linewidth]{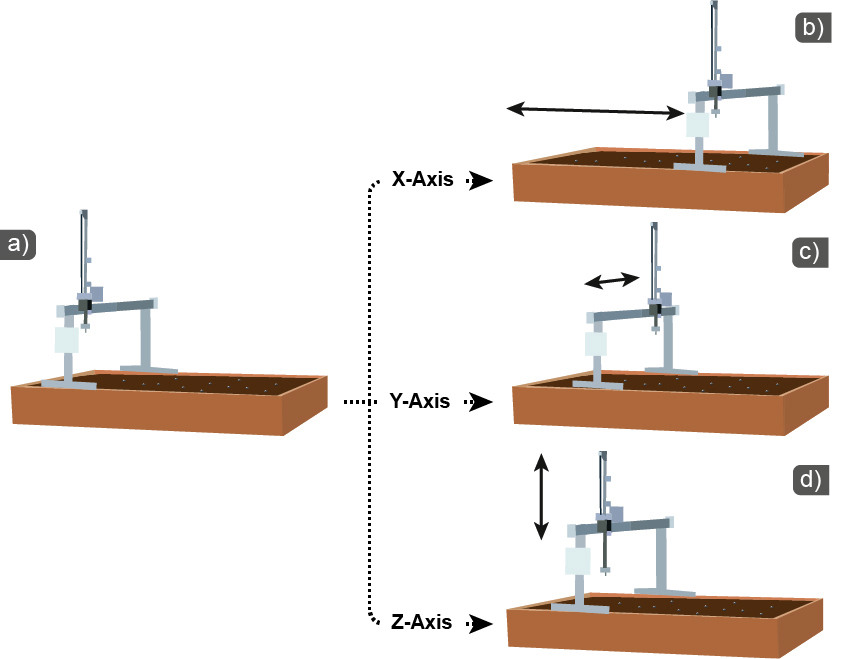}
    \caption{FarmBot moves on a track-based setup as shown in a). This allows for movement across three dimensions: X-axis movement in b), Y-axis movement in c), and Z-axis movement in d).}
    \Description{This image shows the different ways in which FarmBot can move. a) shows the starting position, whereas b) shows X-movement, c) shows Y-axis movement, and d) shows Z-axis movement.}
    \label{fig:movement}
\end{figure}

FarmBot operates on a track-based platform similar to a standard CNC device, enabling movement across the X, Y, and Z axes (\autoref{fig:movement}) with four NEMA 17 stepper motors~\cite{FarmBotTools}. These motors, in combination with a belt pulley system, convert rotational motion into precise linear movement, allowing the robot to navigate the garden bed accurately for tasks like planting, watering, and weeding. Typical of platforms using track-based movement, the area the robot operates within is mapped as a Cartesian coordinate system, with movements specified as three-dimensional coordinates.

\subsubsection{Execution of Tasks}\label{subsubsec:execution-of-tasks}
FarmBot performs tasks as sequences, utilizing five specialized tools for watering, seeding, weeding, and sensing~\cite{BotGeneral}. The Z-axis head of the robot is equipped with a universal tool mount (UTM) featuring twelve electrical connections and magnets. The tools are stored at a fixed location in the field, where a tool bay is installed (\autoref{fig:cams}). The tool pickup process is the same for all gardening tasks.

\begin{figure}[t]
    \centering
    \includegraphics[width=1\linewidth]{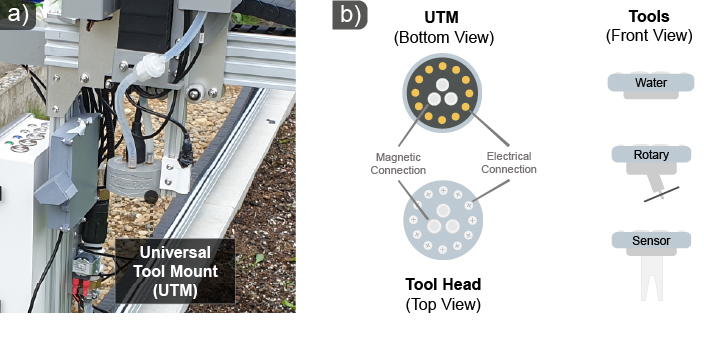}
    \caption{The universal tool mount (UTM) component on FarmBot (a) allows to establish an electrical and magnetic connection to a variety of tools that can be used to execute gardening tasks (b))}
    \Description{This image shows the universal tool mount (UTM) on FarmBot (a), which establishes an electrical and magnetic connection to various tools (b) that enable the performance of diverse gardening tasks.}
    \label{fig:execution-of-tasks}
\end{figure}

The robot moves to the tool's position, lowers its tool mount to connect via magnets, and establishes an electrical connection through the pins (\autoref{fig:execution-of-tasks}). For example, to execute a watering task, the robot first retrieves the watering nozzle by moving to its location, mounting it, and then moving to the designated location to disperse water. The electrical connection between the tool head and the mount enables more complex tasks, such as seeding and weeding. In the case of seeding, the tool head consists of a needle connected to a vacuum pump. After mounting the seeding head, movement to the seed container is initiated. The tool is then lowered while activating the vacuum pump to capture a seed that is transported to the designated planting spot. A rotary motor is used to cut weeds and trim overgrown or weak crop parts.

\subsubsection{Built-in Sensors}\label{subsubsec:built-in-sensing} To enable continuous monitoring of the garden bed, FarmBot collects input through two key sensors. First, one of the tool heads includes a soil moisture sensor~\cite{FarmBotTools}, which provides data to visualize the current moisture saturation. This can be used to adjust watering schedules, suspending them during rain if soil moisture is sufficient. Second, for scanning, mapping, and visualizing the current growth status of the garden, FarmBot uses a USB borescope camera mounted on the Z-axis next to the tool mount (\autoref{fig:cams}). The camera captures images to assess field depth and optimize Z-axis movements for tasks such as seeding, where accurate soil height is essential. Furthermore, these images provide a static visual representation of the current condition of the garden bed.

\subsubsection{Extended Multi-Camera System}\label{subsubsec:multi-cam-system} The built-in borescope camera enables basic plant monitoring but captures single moments, lacking continuous traceability of the robot's actions. This limitation restricts users' ability to fully understand the robot's activities in the field, particularly in a remote gardening setting, where continuous monitoring is critical to maintaining an understanding of ongoing robot actions~\cite{vasconez2019} or encounters with non-human actors~\cite{CameraTrap}. To address this, we extended FarmBot’s camera system with a customized multi-camera setup featuring three additional cameras, bringing the total to four. Each camera is built around a Raspberry Pi 4B\footnote{https://www.raspberrypi.com/products/raspberry-pi-4-model-b Accessed: 24/01/2025} with an attached camera module, positioned to maximize coverage (\autoref{fig:cams}).
Camera A, equipped with a 180° fish-eye lens, was placed at a distance to give an overview of the robot and its position within the field, showing its overall operation. Camera B was positioned on the tool bay's side to capture the robot's home position, allowing users to verify tool pickups and understand the tool attachment process during tasks. Camera C, identical to Camera A and located at the center of the Y-axis, used a 180° fish-eye lens to provide a detailed view of the area, compensating for the borescope camera's limited coverage. This multi-camera system provides a comprehensive overview of plant growth, the robot’s actions, and the spatial context of the field. We designed 3d printable waterproof cases to fit the mounting positions on the robot. The FarmBot's two additional 24V pins were used as power supplies for our camera system. The camera streams were made accessible online via a web server (Ubuntu 20.04 LTS), running a MediaMTX media proxy\footnote{https://github.com/bluenviron/mediamtx Accessed: 24/01/2025}.
\begin{figure*}[t]
    \centering
    \includegraphics[width=1\linewidth]{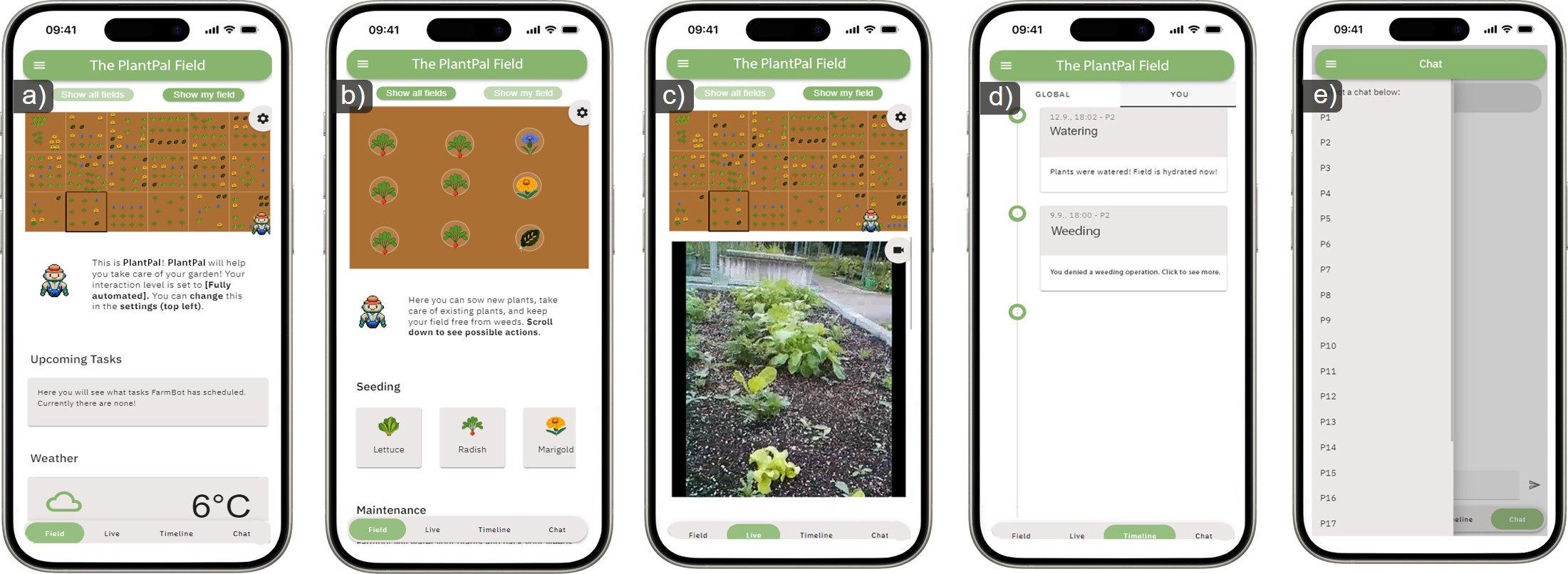}
    \caption{An overview of the \system web application. It is optimized for mobile use to make it accessible on demand. a)The global field view provides an overview of all fields on the \system field. Users can use scrolling and dragging gestures to zoom into other gardeners' plots and review what they have planted. b) Similarly, a personal field view is provided that is focused on the execution of gardening tasks, designing a garden layout, and reviewing progress. c) At any time, users can active a live stream that enables real-time monitoring from three different perspectives. Displaying the live stream at the bottom allows for parallelization between virtual and real-world aspects. d) Using the timeline, users can review their own decisions or, if they chose a higher automation level, the action that the PAR performed during their absence. e) Lastly, a chat was implemented as a way for conflict resolution between gardeners and to support social interaction more broadly.}
    \Description{Figure 5 shows an overview of the plantpal application. a) shows the global field view provides an overview of all fields on the \system field. Users can use scrolling and dragging gestures to zoom into other gardeners' plots and review what they have planted b) shows a personal field view is provided that is focused on the execution of gardening tasks, designing a garden layout, and reviewing progress c) At any time, users can active a live stream that enables real-time monitoring from three different perspectives. Displaying the live stream at the bottom allows for parallelization between virtual and real word aspects d) Using the timeline users can review their own decisions or if they chose a higher automation level, the action that the PAR performed during their absence e) Lastly, a chat was implemented as a way for conflict resolution between gardeners and to support social interaction more broadly.}
    \label{fig:overview}
\end{figure*}

\subsection{The \system Web Application} \label{subsec:app-overview}
To control the FarmBot, a fully open-source web application is already provided by the developers\footnote{https://github.com/FarmBot/Farmbot-Web-App Accessed: 24/01/2025}. With it, users can fine-tune settings, create custom sequences/routines, and obtain visualizations of their gardens. Like most CNC control applications, the integrated functions favor tech-savvy users interested in fine-tuning the system and exploring its functionalities. Further, FarmBot's innate web application does not foresee multiple users' shared use of one robot. Instead, it is designed to provide control to one nearby user who owns the robot. Inspired by the existent web application, we sought to implement a customized version that enables the shared use of one FarmBot. Additionally, this enabled us to align control mechanisms, visualizations, and field design mechanisms with the design goals outlined in \autoref{subsubsec:design-goals}. Our full-stack web application, \system, was 
implemented using Nuxt3\footnote{https://nuxt.com/ Accessed: 24/01/2025} and VueJS\footnote{https://vuejs.org/ Accessed: 24/01/2025}. VueJS was used for frontend development, while Nuxt3 was used for backend development. To ensure reliable management of multiple users and logging of user actions, we used a MySQL database. To communicate with FarmBot, \system further leverages the FarmBotJS and OpenFarm API~\cite{FarmBotDocs}. Interacting remotely with a garden plot by collaborating with a PAR allows for always-accessible and on-demand engagement with urban gardening. Further aligning with \textit{D1}, \system's design was optimized for mobile devices to ensure that access to the PAR and the multi-camera system can be achieved at any point throughout the day.  

\subsubsection{General Overview}\label{subsubsec:app-general-overview}
To use \system, each user receives an individual user account with personalized log-in information. Upon login, the user is first given a general overview of the layout and functionalities. As shown in \autoref{fig:overview}, \system featured a two-row layout where the top row represents a map view of the field and the bottom row provides UI elements that consist of additional relevant information such as the weather at the field's location, daily plant care tasks, plant care actions (e.g., weeding, seeding, and watering), and access to the multi-camera live stream. The top row can assume two states: a (zoomable) full-field overview or an already zoomed-in field view focusing on one's personal plot. The following section elaborates on strategies developed to address the established design goals.

\begin{figure*}[t]
    \centering
    \includegraphics[width=1\linewidth]{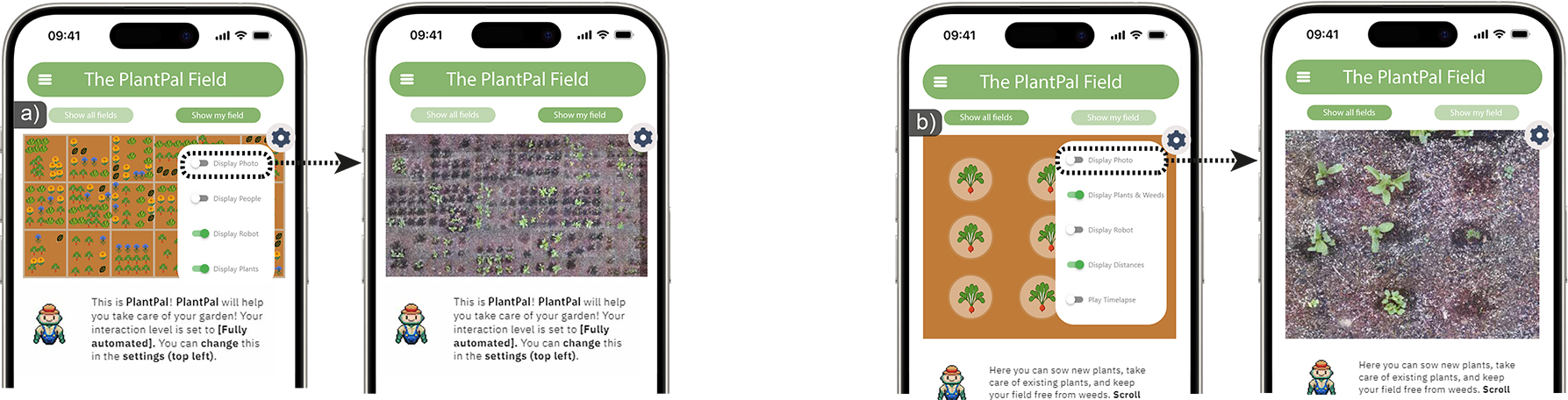}
    \caption{The degree of detail and realism regarding field visualizations can be dynamically adapted. a) shows how a photo grid overlay can be visualized for the global field. Similarly, b) showcases this for a personal field plot.}
    
    \Description{This image shows how the degree of detail and realism in field visualizations can be dynamically adapted. (a) illustrates a photo grid overlay for the global field, while (b) showcases the same for a personal field plot. In both views, you can see the actual field progress, with green parts representing areas where plants are growing.}
    \label{fig:map-viz}
\end{figure*}

\subsubsection{Control Strategies}\label{subsubsec:control-strategies}
FarmBot is typically configured such that users can set timed routines for watering, seeding, and weeding tasks. However, customized control strategies were required in our multi-user, remote gardening approach to offer flexibility between more active and passive control. Leaning on existent robot control strategies (e.g.,~\cite{vasconez2019, pang2014study, endsley}), we implemented three control modes: (1) \textit{Manual}, (2) \textit{Hybrid}, and (3) \textit{Automated}. Each mode balances initiative in decision-making differently between the user and the robot.
Using \textit{Manual} mode, users control all key decisions, such as where to sow seeds, water, or remove weeds, with the robot merely executing tasks and offering \textit{non-binding} warnings, such as when seeds are sown too close together. 
In contrast, \textit{Automated} mode permits FarmBot to manage \textit{all tasks} autonomously, with users serving as spectators except during planting, where users can choose the crops they would like to plant. At the same time, the location-based algorithms determine the optimal position instead of the user. 
In \textit{Hybrid} mode, the robot takes a more active role in decision-making. While users still guide tasks, such as selecting plant types, the robot intervenes to prevent mistakes (\textit{i.e. non-binding warnings turn into binding restrictions going from \textit{Manual} to \textit{Hybrid}}). For instance, it will stop the user from planting seeds too close together or watering excessively. This mode requires less fine-grained decision-making from the user, with the robot ensuring that critical gardening errors are avoided while still allowing the user to know when and if tasks should be executed. 
\system allows users to switch between these modes based on their willingness or ability to take initiative in decision-making for tasks related to plant health. \textit{\system does not automatically adapt} the modes based on user profiles or behavior as changing between control modes already adapts the options and restrictions \system provides. The flexible switching mechanism aligns with the aims outlined in \textit{D2} as the user can freely choose their level of involvement.

\subsubsection{Digital Augmentations of the Gardening Process}\label{subsubsec:digital-augmentation}
\system incorporates a variety of visualizations designed to augment the remote gardening experience to provide alternative perspectives unique to the remote setting. While \system allows users to trigger actions on garden plots from anywhere remotely, our formative survey revealed concerns about establishing a personal connection with the plants and field. Drawing inspiration from recent work on technology-supported nature engagement, highlighting that distant interaction can offer alternative experiences distinct from in-person engagement~\cite{Webber2023CHI,mediatedNature,generatedNature}, we designed \system to embrace this notion, especially in connection to \textit{D3}.

\system provides several digital augmentations. One feature is the daily capture of field images by FarmBot, offering users concurrent insight into the current state of their garden. These images are overlaid with practical information, such as the exact locations of newly planted seeds, even when only soil is visible until germination. Leaning on Farmbot's innate web application~\cite{FarmBotDocs}, we visualize expected plant growth through dynamic growth circles to offer a predictive view of how large each crop will grow. This allows users to manage present tasks and anticipate future developments.

Additionally, daily images are saved and compiled into a time-lapse, which extends as the user interacts with their field over time. Using the field visualization, users can play this time-lapse at any point. Building on prior research (e.g.,~\cite{reflection1, mediatedNature}), this feature aims to encourage reflection on past developments leading to the present state. In-person gardening, by contrast, typically focuses on the immediate present as gardeners physically interact with their plants in real-time, responding to visible needs such as watering or weeding as they arise. Users can toggle between abstract views and detailed real-time representations of their garden. Inspired by FarmBot's web interface, \system’s switching mechanism lets users choose their preferred degree of detail between data-rich or abstract representations. We argue that this approach aligns with the crop cultivation process, as gardeners may require detailed information for decision-making when plant care actions are executed, while at other times, they may only need to check on the field to verify its current state without needing comprehensive data. This flexibility is designed to enable seamless personalization of the user experience when interacting with \system.

\subsubsection{Inter-Gardener Relations}\label{subsubsec:inter-gardener-relations}
While for some individuals, gardening is primarily outcome-focused (e.g., harvesting crops)~\cite{garcia2018,opitz2016}, others emphasize the process, which encompasses more than the physical care of plants~\cite{cumboAdverse2020,bidwell2010}. Social interaction, especially in urban community gardens, is a key component of the gardening experience, fostering communication and collaboration among gardeners~\cite{ding2022,gough2013a,fox-kamper2018b}. Our formative survey indicated that participants viewed shared features as important for a system like \system, which facilitates remote engagement with garden spaces. Gardening with \system mirrors aspects of traditional shared gardens, where users manage individual plots in a shared space. In community gardens, collaboration often involves knowledge exchange and coordinated plant care~\cite{knowledgeSharing}. \system, however, introduces a PAR as a \textit{constant gardening partner}, adding a unique dynamic. While the PAR is a shared resource, effective communication remains relevant, especially as users may engage at varying control levels (\textit{Manual, Hybrid, Automated}).
Conflicts may arise when gardeners operate differently. For instance, users in \textit{Manual} mode might plant near plot borders, causing overgrowth into neighboring spaces and impacting others. To resolve such issues, \system includes a chat function for direct communication among gardeners sharing the robot (\autoref{fig:overview}). Additionally, \system implements a First-Come-First-Served task queue to manage access when multiple gardeners request the robot within a short time.
The global field view and event timelines (\autoref{fig:overview}) allow users to monitor each other's progress and the field's overall state. In \system's current implementation, users cannot edit the progress shared with others. This decision prevents scenarios where users hide information, which could distort the visual representation of progress on the global field and potentially lead to demotivation due to excessive hidden data. The above aspects align with \textit{D3}, providing a dimension of engagement beyond gardening tasks.

\section{Evaluation}\label{sec:evaluation}
To explore how users interact with \system and assess the impact of robot-assisted garden cultivation on gardening outcomes, users' connection to their plots, and attitudes toward urban gardening, we conducted a 3-week exploratory field study.
\subsection{Study Design}\label{subsec:evaluation}
Recent research suggests that longitudinal studies investigating the effects of technology-supported interactions with natural environments remain rare~\cite{Webber2023CHI}. We deployed FarmBot on an 18m² field near our institute, dividing it into 18 plots ($1m x 1m$) assigned to participants. In our freestanding setup without protective structures (e.g., greenhouse), we evaluate \system over 3 weeks under real-world conditions.
Participants were told that they could use \system to design a personalized garden layout and cultivate their assigned plots over the study period. The study was conducted at the Botanical Garden of the University of Ulm, Germany, and participants were compensated with €30,00. The study was carried out in full compliance with the ethical guidelines and regulations established by the university's review board.
\begin{figure*}[t]
    \centering
    \includegraphics[width=1\linewidth]{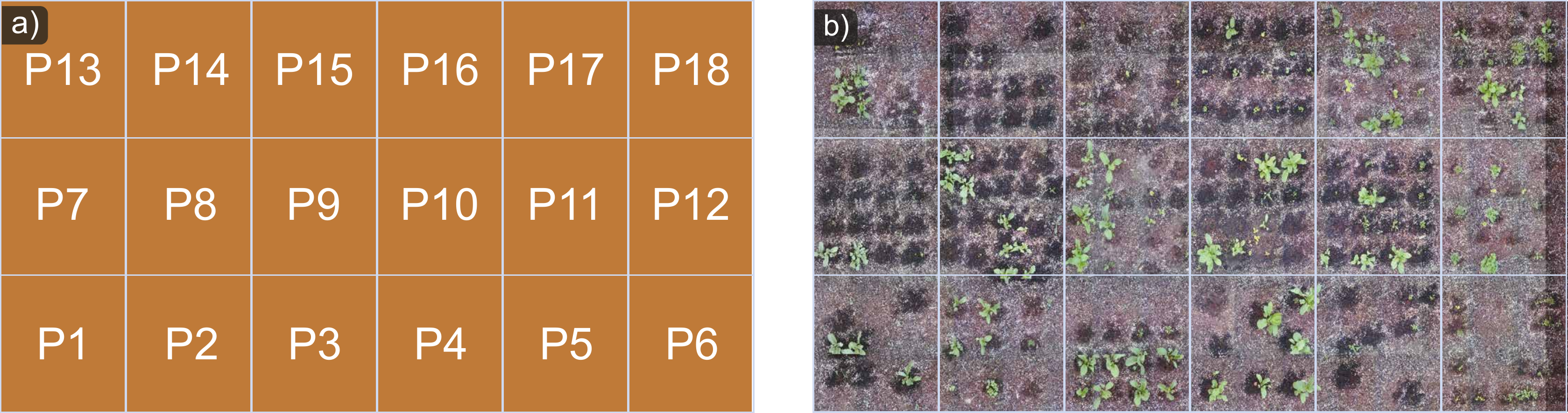}
    \caption{Illustration a) shows the distribution of participants on the virtual field and b) the same distribution on the real garden bed as sampled from the \system app (top-down view).}
    \Description{This figure shows how participants were distributed on the PlantPal field. At the bottom left P1 was assigned. P2-P6 are next to P1, P7 is above P1 starting a new row. This pattern is repeated for 18 participants.}
    \label{fig:participant-distribution}
\end{figure*}
\subsubsection{Measures}\label{subsubsec:metrics-main-eval} 
For quantitative metrics, we assessed participants' connection to their field using the Inclusion of Other in the Self (IOS) Scale~\cite{IOS1,IOS2}. We also used subscales from the Environmental Attitude Inventory (EAI)~\cite{MILFONT201080} to assess views on perceived enjoyment, alteration, conservation, and dominance over nature. These scales assessed shifts in participants' perceptions of technology use in natural environments, their roles, and green self-perception related to cultivating a garden. Our formative survey highlighted enjoyment of and connectedness to nature as concerns (cf. \autoref{subsubsec:dg-qualitative}). These measures enabled us to determine how effectively \system addressed these concerns.
Additionally, we measured overall user experience with \system using the brief version of the User Experience Questionnaire (UEQS)~\cite{UEQS}, usability using the System Usability Scale (SUS)~\cite{SUS}, and included single item questions to capture sentiments about remote gardening and collaboration with PARs. User logs were recorded to track how participants integrated \system into their daily routines and how they used specific features. Finally, we conducted voluntary semi-structured interviews (similar to~\citet{CameraTrap}) to gain deeper insights into participants' strategies and perceptions, as well as to clarify patterns observed in the user logs. The following describes our sample and the study procedure.

\subsubsection{Participants}\label{subsubsec:participants-main-eval} 
\begin{figure*}[t]
    \centering
    \includegraphics[width=1\linewidth]{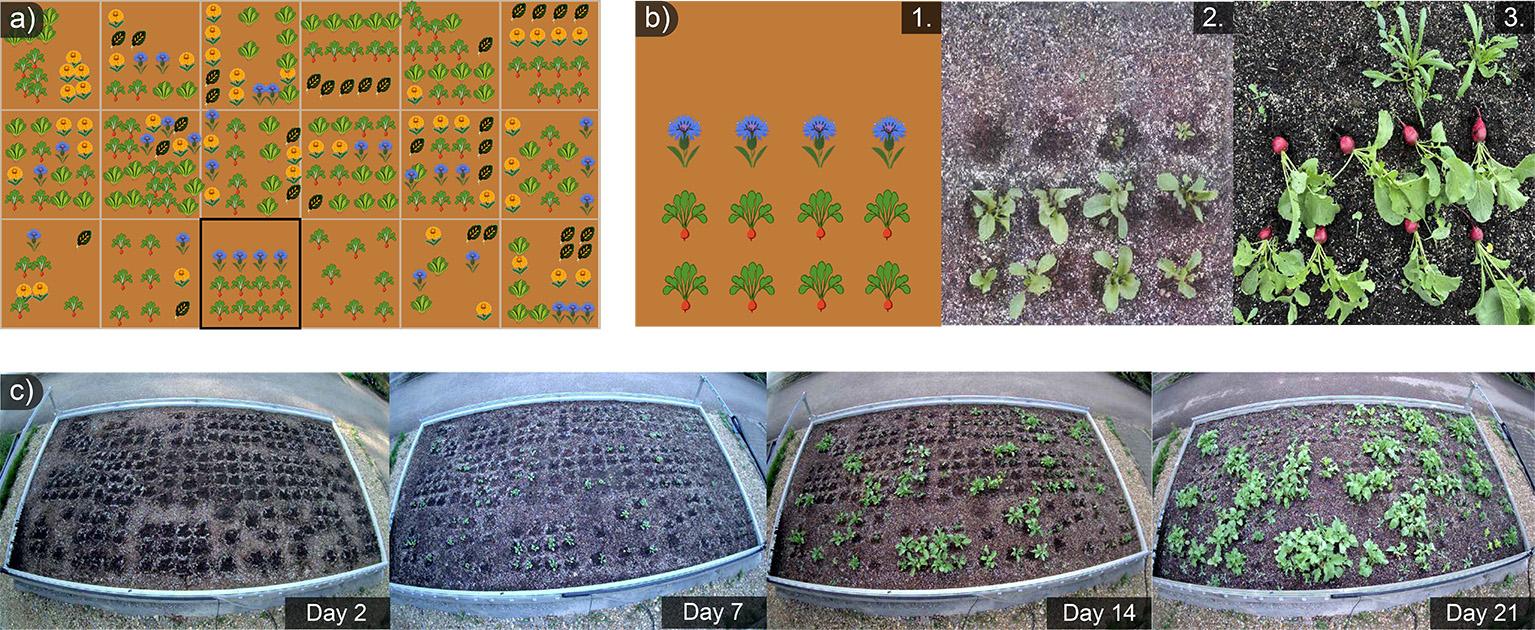}
    \caption{Participants planted various crops in their fields. The resulting distribution of crops on the entire field is shown in a). \system allowed the participants to focus on their personal field view in addition to the global view. b) shows the example of P3, who planted two rows of radishes and one row of cornflowers (b.1). This virtual crop distribution was replicated by FarmBot on a real garden plot (b.2) and led to successful cultivation toward the conclusion of the study (b.3). Similarly, the progress across selected time points during the 3-week study can be seen in c). The precise and continuous watering within the same locations that were specified virtually by participants led to soil displacements that make the distribution of crops seen in a) visible on the real garden bed in c). The maturity of most crops was not reached at the conclusion of the study, yet successful cultivation across all fields can be observed (c-Day 21)}
    \Description{This image shows the planting and cultivation process during the study. (a) illustrates the distribution of crops planted by participants across the entire field. The system allowed participants to focus on their personal field view in addition to the global view. (b) provides an example of P3's personal field: (b.1) shows two rows of radishes and one row of cornflowers as planned virtually, (b.2) depicts this distribution replicated by FarmBot on a real garden plot, and (b.3) highlights successful cultivation toward the end of the study. Similarly, (c) shows the progress across selected time points during the 3-week study. Green parts indicate crop growth, and the precise, continuous watering at the virtual locations specified by participants caused soil displacements that made the distribution of crops in (a) visible on the real garden bed in (c). Although most crops had not reached full maturity by the study's conclusion, successful cultivation is evident across all fields (c-Day 21).}
    \label{fig:garden-results}
\end{figure*}
Initially, we recruited participants from our personal networks. We further used snowball sampling~\cite{goodman1961snowball} to gather a total set of 18 participants. Participants were required to be individuals whose ability to engage in gardening \textit{to their satisfaction} is hindered by one or more barriers outlined in \autoref{subsec:rel-challenges}. The number of participants matched the available plots in our field (\autoref{fig:participant-distribution}). Participants' ages ranged from 21 to 64 ($M=33.33$, $SD=14.84$). Ten participants identified as female and eight as male. Seven participants held a bachelor's degree, six held a high school diploma, three held a master's degree, one held a Doctorate, and one had completed an apprenticeship. Eight participants were students, six were employed, two were out of work, and two were self-employed. Regarding living situations, 12 participants lived in private apartments without access to a backyard garden, four lived in houses with shared gardens suitable for cultivating crops, and two resided near green spaces where crop cultivation was not permitted. We included four participants with access to a gardening space. These participants expressed interest in utilizing \system to grow crops for which they lacked space in their current gardens, replicating the concept of allotment gardening. Including these participants allowed us to explore how individuals with existing green space but spatial limitations would engage with \system. Sixteen participants had tried cultivating plants before participating in our study, while two were novices with no prior gardening knowledge but expressed a strong interest in learning. Based on the Affinity for Technology Interaction (ATI) Scale~\cite{ATI1,ATI2}, participants scored $M=3.92$ $(SD=0.93)$, reflecting moderate familiarity and interest in digital systems.

\subsubsection{Procedure}\label{subsubsec:procedure-main-eval} 
Each participant went through a kick-off session where they were introduced to the study process, provided informed consent, and completed a pre-study questionnaire. In the introduction, participants received personalized login information for the \system web application and were guided through the functionalities. The layout of \system was described to the participants, and they were further shown how the individual control modes (i.e., \textit{Manual, Hybrid, Automated}) differ (cf. \autoref{subsubsec:control-strategies}). Lastly, participants received a brief walkthrough on adding crops to their fields, watering them, and conducting weed management via the \system interface. They were further made aware that progress on the app would be visible to other participants. As FarmBot requires seeds for desired crops to be supplied in a container beforehand, we curated a selection of crops based on participant feedback during the recruiting process. With the selection we offered, we aimed to address different gardening motivations such as food production~\cite{opitz2016, garcia2018}, support for biodiversity~\cite{tresch2019,clucas2018}, or aesthetic appeal~\cite{murtagh2023}. The resulting list of crops that could be cultivated during the study consisted of lettuce, radish, cornflower, marigold, and cumin. Before the study, the chosen seeds were sorted and laid out in FarmBot's seeding containers to make them accessible. Apart from setting up FarmBot itself, no other on-site intervention was required. Most participants then started designing their personal field layouts towards the end of the kick-off session and used the system for three weeks. During the evaluation period, participants could contact the study supervisors via a contact form embedded within the \system web application for any questions. At the end of the three weeks, participants completed a post-study questionnaire and could participate in voluntary semi-structured interviews. Afterward, participants were compensated and asked if they would like to keep their \system access for the remainder of the growing season.

\section{Results}
\subsection{Garden Cultivation}\label{subsec:results-garden-cultivation}
\autoref{fig:garden-results}c shows the participants' individual garden plots at different time points throughout the study. Participants took different approaches to designing their field plot layouts. Thirteen participants manually placed crops, while five selected the crops but let FarmBot arrange them, leading to machine-like grid layouts. 250 crops were initially planted, with an average of $M=13.89$ $(SD=3.92)$ per field. Seventeen participants chose a mix of crops, while one participant exclusively cultivated radishes on their plot (P4). Planting success was gauged based on the percentage of crops that reached germination and progressed beyond early growth stages. This was judged visually and using the FarmBot's plant growth monitoring. On average, $M=80.75\%$ $(SD=17.18\%)$ of the seeds planted successfully grew past the germination stage. Using the log files gathered during the three weeks, we observed preferences for one of the three control modes (cf. \autoref{fig:switches}). We define such a preference as more than 50\% of the study duration spent in one mode. In particular, \textit{P5, P7, P9, P10, and P17} spent most of the study duration in \textit{Automated}, \textit{P1, P12, and P15} leveraged \textit{Manual} the most, and  \textit{P2-P4, P6, P8, P11, P13, P14, P16, P18} preferred using \system in \textit{Hybrid}. Participants preferring \textit{Automation} mode successfully cultivated $M=75.00\%$ $(SD=15.81\%)$, those who preferred \textit{Manual} mode successfully cultivated $M=75.00\%$ $(SD=20.41\%)$, and participants preferring \textit{Hybrid} mode successfully cultivated $M=85.35\%$ $(SD=14.29\%)$.

\subsection{Questionnaire Data}\label{subsec:results-questionnaire-data}
We assessed normality using Shapiro-Wilk tests to determine the appropriate statistical method to examine differences between pre- and post-study ratings.
\begin{figure}
    \centering
    \includegraphics[width=1\linewidth]{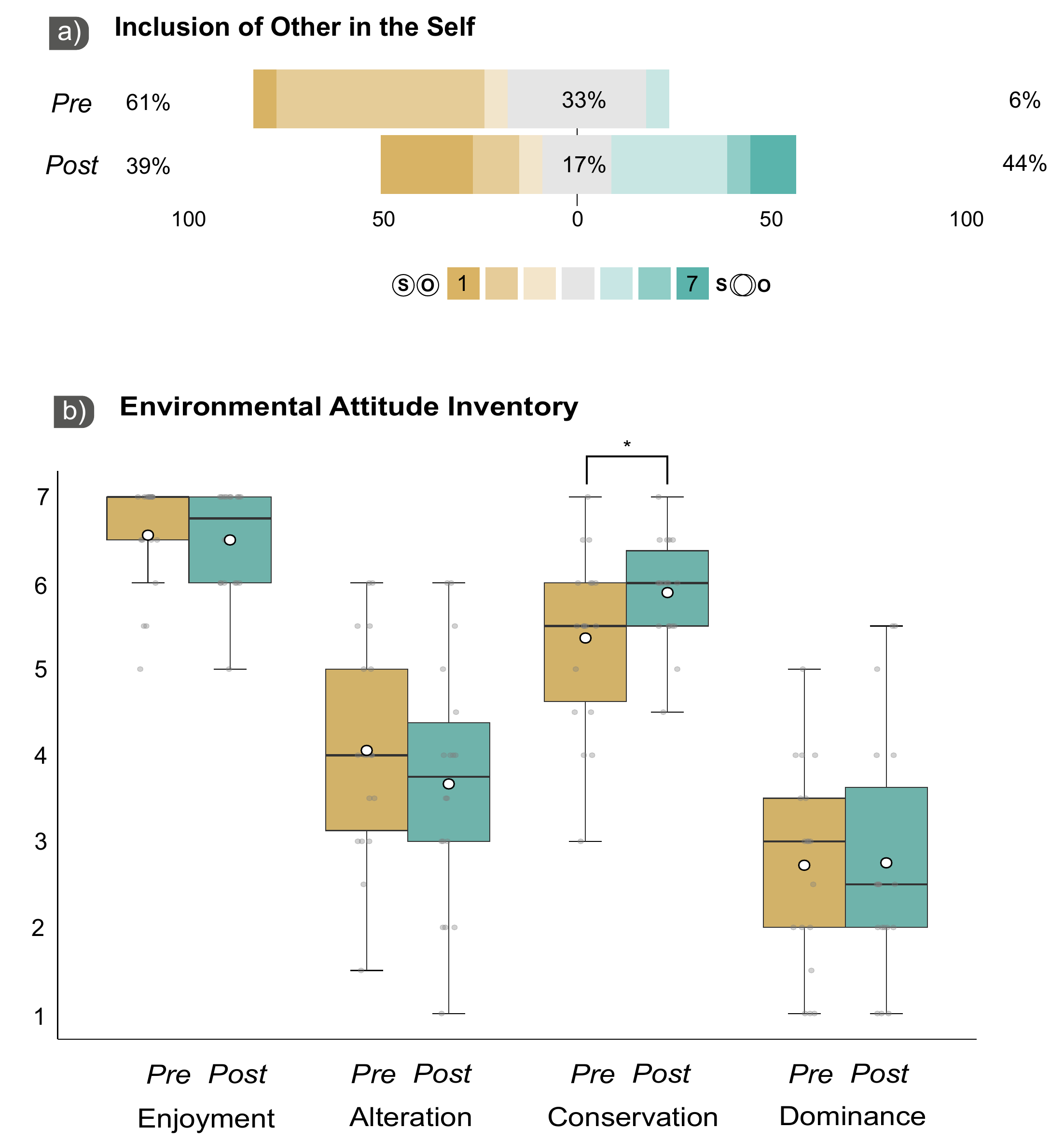}
    \caption{Distributions for a) Inclusion of Other in the Self (IOS) Scale and b) average scores for the subscales \textit{Enjoyment of Nature, Altering Nature, Dominance over Nature, and Personal Conservation Perception} in the Environmental Attitude Inventory (EAI). Pre- and post-study results are shown. The significant increase ($p<0.05$) for the conservation subscale is denoted with *.}
    
    \Description{The figure shows the distributions of two types of data collected before and after the study: a) The Inclusion of Other in the Self (IOS) Scale, which measures participants' sense of closeness or connection with nature. b) Average scores for four subscales from the Environmental Attitude Inventory (EAI): Enjoyment of Nature, Altering Nature, Dominance over Nature, and Personal Conservation Perception. Pre- and post-study comparisons reveal a significant increase in participants' connectedness to nature (IOS) and their conservation-oriented attitudes (Personal Conservation Perception), indicating a stronger relationship with and concern for the environment over time.}
    \label{fig:ios-eai}
\end{figure}

\subsubsection{EAI}\label{subsubsec:results-eai}
Dominance over nature, the appropriateness of altering nature through human intervention, enjoyment of nature, and perception of one's conservation behavior were measured via EAI subscales (7-point Likert ratings). Paired t-tests revealed no significant differences for dominance, alteration, and enjoyment of nature subscales. However, a paired t-test on the subscale of personal conservation behavior showed a significant increase between pre- ($M= 5.36, SD=1.03$) and post-study ($M= 5.89, SD= 0.61$) ratings ($t(17)=-2.82, p=0.012, d=0.66$) (\autoref{fig:ios-eai}). 

\subsubsection{IOS}\label{subsubsec:results-ios}
A paired t-test indicated a significant increase between pre- ($M=2.83, SD=1.15$) and post-study ($M=4.06, SD=1.98$) ratings ($t(17)=-2.15, p=0.046, d=0.51$). Additionally, we descriptively compared pre- and post-study ratings grouped by preferred control mode. No statistical tests were conducted due to the small and uneven group sizes. IOS scores increased by $0.2$ (pre: $M= 2.2, SD= 1.1$; post: $M=2.4, SD= 1.52$) for participants preferring \textit{Automated} mode. For those preferring \textit{Manual} mode, IOS scores remained the same (pre: $M= 4.0, SD= 1.0$; post: $M=4.0, SD= 1.73$) and increased by $1.6$ (pre: $M= 2.8, SD= 1.03$; post: $M=4.4, SD= 2.17$) for those preferring \textit{Hybrid} mode.

\subsubsection{SUS \& UEQS}\label{subsubsec:results-ueq}
The resulting average score of $M=78.38$ $(SD=10.82)$ on the SUS indicated an above-average level of usability, according to~\citet{SUS}. Evaluation of the UEQS resulted in a score of $M=1.38$ $(SD=1.09)$ on the pragmatic quality subscale, $M=1.54$ $(SD=0.68)$ for hedonic quality, and an overall score of $M=1.46$ $(SD=0.61)$, indicating a positive user experience, with the system being perceived as functional and pleasant~\cite{UEQS}.

\subsubsection{Single-Item Ratings}\label{subsubsec:results-single} Single-item Likert ratings were used to measure changes in motivations for gardening, the willingness to interact with robotic actors to garden, and preferences for personal involvement in gardening. A paired t-test yielded significant differences for the statement \textit{"I perceive gardening as an activity I would do primarily to grow my own food."} between pre- ($M= 5.72, SD=1.32$) and post-study ($M= 4.89, SD= 1.53$) ratings ($t(17)=-0.53, p=0.04, d=-0.50$). Further, regarding willingness to interact with a robot for collaboration in urban gardening, an exact Wilcoxon-Pratt Signed-Rank test found a significant difference between pre- ($M=1.61$, $SD=1.04$) and post-study ($M=2.33$, $SD=1.57$) ratings for the statement \textit{"I would use smart devices such as farming robots for gardening."} ($Z=-2.44, p=0.031, r=0.65$). No further significant differences between pre- and post-study answers were found for the remaining ratings. 

\subsubsection{Retention}\label{subsubsec:results-retention} Since three weeks were insufficient for crops to mature fully, participants were offered extended access to \system and their field. Sixteen participants agreed: nine to harvest their crops, five to donate them, and two were motivated primarily by satisfaction. Participants who did not wish to extend their access to \system mentioned digital detox as a reason (P13) and felt that using \system on a laptop would be preferable for them (P10). However, \system was optimized for mobile specifically.

\begin{figure}[t]
    \centering
    \includegraphics[width=1\linewidth]{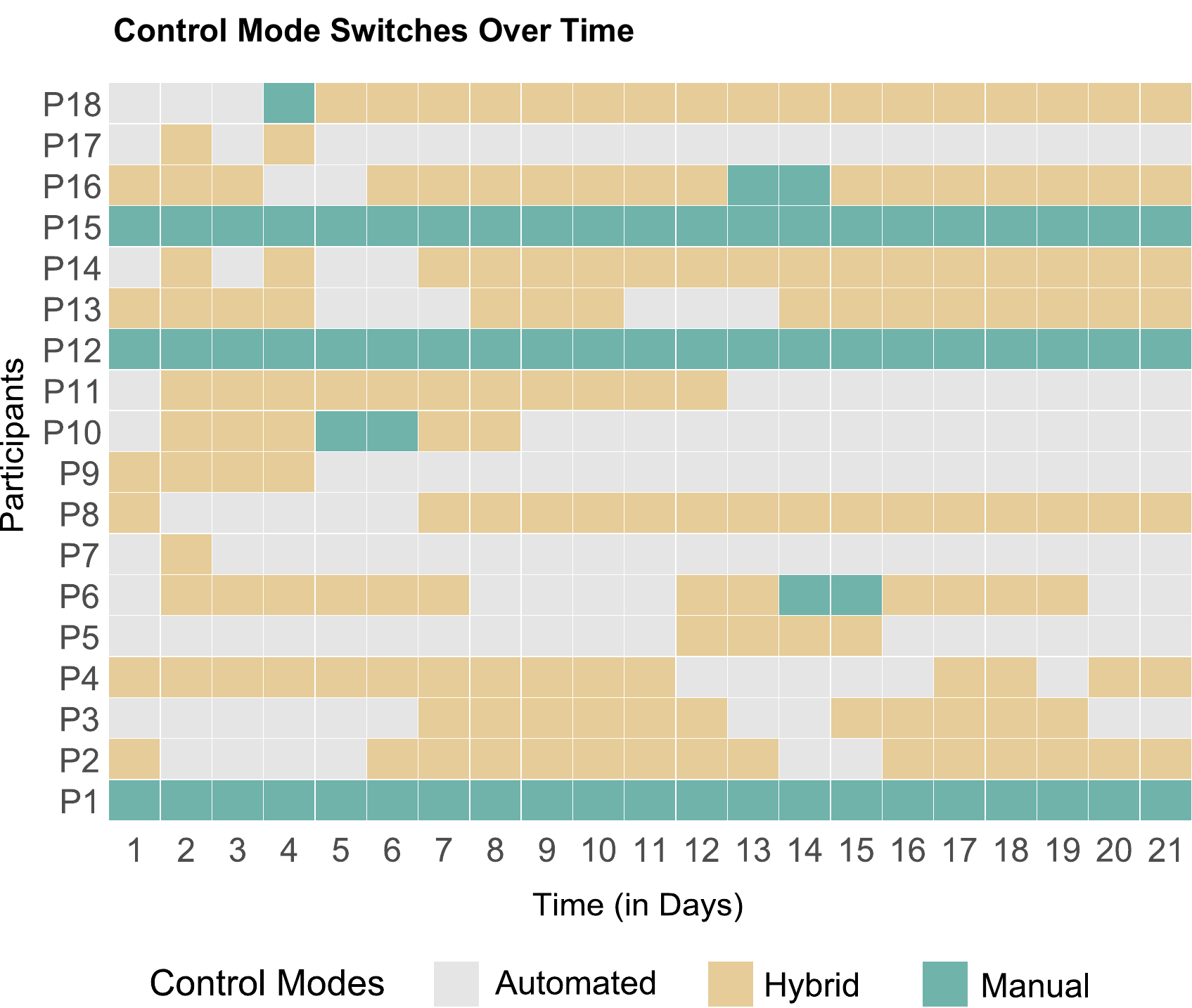}
    \caption{A scarf plot showing the control modes participants used throughout the study duration (grey=automated, yellow=hybrid, and green=manual). The control mode used the longest on a given day was estimated based on user logs to assign the visualized labels.}
    \Description{This Figure shows how users switched between different control modes. It can be seen that some participants remained consistent with their control mode like P1, P12, and P15. Others switched quite frequently, especially in the beginning.}
    \label{fig:switches}
\end{figure}
\begin{figure}[ht]
    \centering
    \includegraphics[width=1\linewidth]{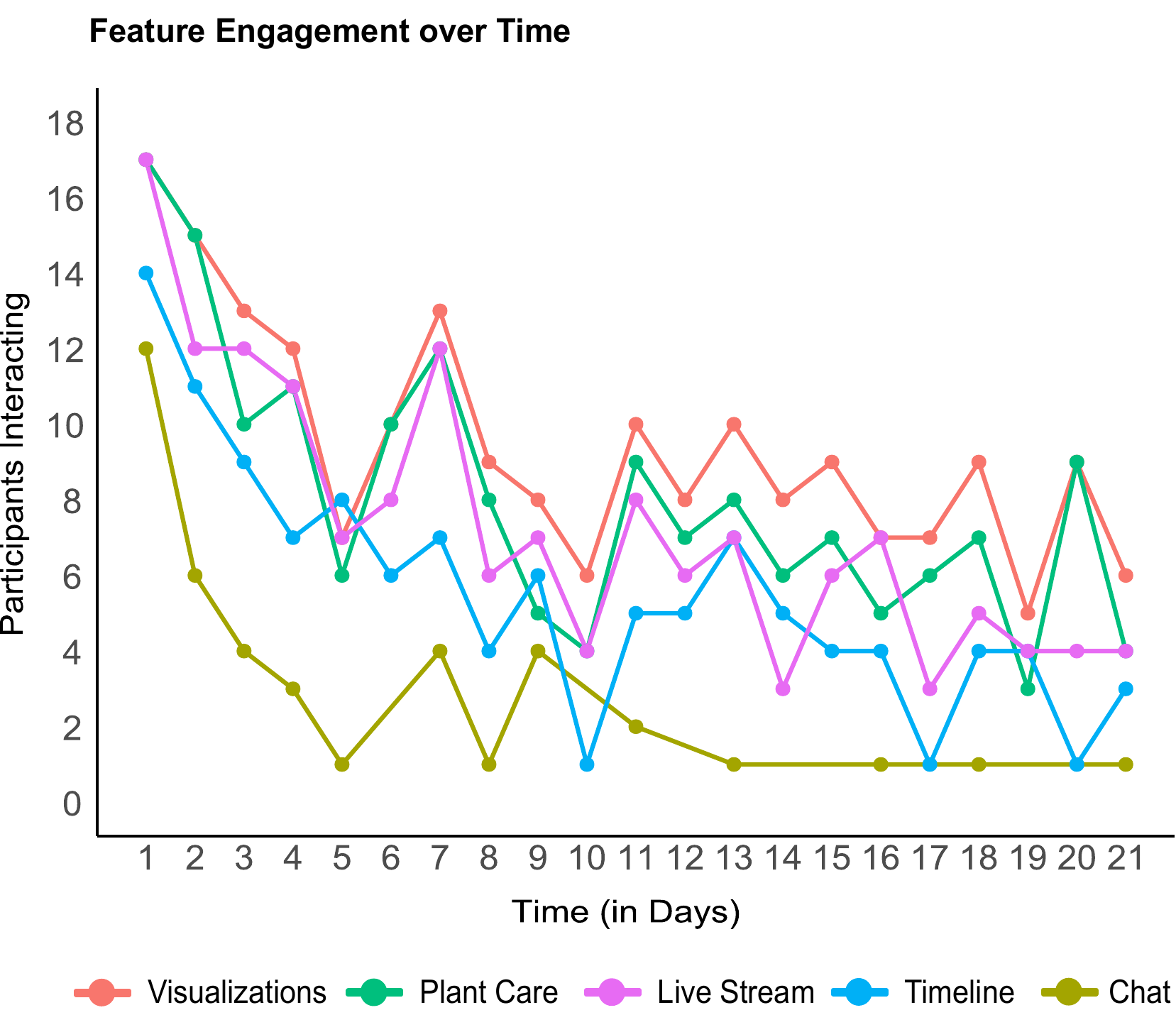}
    \caption{Line plot illustrating how users engaged with \system over the study period. We visualized user logs by clustering actions according to their goal (visualization, plant care, live stream, timeline, and chat). \textit{Participants Interacting} represents the number of unique participant IDs counted for the feature categories for each day.}
    \Description{A line plot showing user engagement with \system throughout the study period. The visualization clusters user actions into five categories: visualization, plant care, live stream, timeline, and chat. The metric "Participants Interacting" indicates the daily count of unique participant IDs engaging with each feature category. The plot reveals a drop in engagement after the first week, followed by stabilization during weeks two and three.}
    \label{fig:user-actions}
\end{figure}
\subsection{User Logs}

\subsubsection{Login Behavior}\label{subsubsec:results-logins}

We recorded a total of \textit{1217} logins into \system, where logins were defined as opening the \system web page, while logoffs could either be counted as switching to a different tab, closing the tab, or closing the browser. Login durations varied ($M=4.16mins$, $SD=20.26mins$) and occurred distributed throughout the day. Longer login durations may have been due to participants not explicitly closing the \system window or browser. Applying outlier filtering results in a login time of $M=1.31mins$ $(SD=1.35mins)$. The majority of users (37.72\%, 459 logins) were active after 4 PM. The remaining logins occurred either in the morning or throughout the day. On average, 14 participants ($M=14.14$, $SD=3.02$) logged in daily in the first week, 9 in the second week ($M=9.14$, $SD=1.21$), and 9 in the third week ($M=9.00$, $SD=1.15$). All participants logged in at least once a day in the first week. In the second week, 16 participants logged in at least once a day, and 17 in the third week.

\subsubsection{Actions \& Patterns}\label{subsubsec:results-actions-and-patterns}
Participants performed various actions, such as examining map visualizations (\textit{33.39\%}), executing gardening tasks (\textit{13.48\%}), viewing one of the live streams (\textit{11.79\%}), reviewing the robot's actions using the (personal and global) timeline (\textit{7.81\%}). The chat feature was used the least (\textit{1.66\% of all actions, used only by P8 and P13}). Participants did not always explicitly change map filters and instead merely switched between global and personal field windows (\textit{30.47\%} of actions). We additionally analyzed which features were often used in conjunction. This was done by grouping actions as tuples based on timestamps (i.e., actions occurring right after another). Here, it became apparent that logging in to execute a gardening task was frequently followed by reviewing the live streams. Further, participants tended to review the timeline followed by the current image of one's field, resembling brief "checkups". 

\subsubsection{Longitudinal Changes in Behavior}\label{subsubsec:results-longitudinal-changes}
Control mode logs showed preferences for specific modes, yet, on average, participants switched between control modes $M=2.8$ $(SD=1.75)$ times. Control mode switches occurred throughout the study period (cf. \autoref{fig:switches}), with some users (e.g., P1, P12, P15) not switching at all or throughout the later portions of the study duration (e.g., P4, P6). While 14 participants did at some point try the \textit{Automated} mode, the fully \textit{Manual} mode was either only used shortly or throughout the entire study duration. Participants frequently switched between \textit{Automated} and \textit{Hybrid}.

\autoref{fig:user-actions}b illustrates the total number of participants performing different actions on \system. Plant care actions by automated users, as well as simple logins without UI interactions, were excluded from tracking. All participants initially performed plant care actions and engaged with visualization and surveillance features (e.g., map views and live streams) during the sowing phase. However, daily plant care actions decreased after the first week. A similar number of participants utilized visualization ($M=12.43$, $SD=3.26$) and live stream features ($M=11.29$, $SD=3.25$) during the first week.  Live stream usage declined over time, while visualizations experienced similar use. Plant care actions became more sporadic towards the study's conclusion. On average, the timeline feature was used by eight participants in the first week ($M=8.86$, $SD=2.79$), five in the second ($M=4.71$, $SD=1.89$), and three in the third ($M=3.00$, $SD=1.41$) week. While multiple participants opened the chat, only two actively sent messages. The average number of actively interacting participants (i.e., those not only logging in to check but triggering UI-based interaction events regardless of the category) in the first week was 14 ($M=13.57$, $SD=2.88$), 9 in the second ($M=8.57$, $SD=1.40$), and 8 in the third week ($M=8.00$, $SD=1.83$).

\subsection{Qualitative Data}\label{subsec:qualitative data}
\subsubsection{Analysis}\label{subsubsec:TA-analysis}
Out of 18 participants, 12 (P1-P7, P10-P12, P15, and P18) participated in voluntary post-study interviews. Interviews lasted $M=24.07mins$ ($SD=8.55mins$). We defined a set of questions (see \autoref{appendix:A}) for the semi-structured interviews. Further, for each participant, we also reviewed the respective user logs before the interview sessions and noted interesting or irregular behaviors to gain more context from the participants. The interview audio files were first transcribed using whisperX~\cite{bain2023whisperx}. Moreover, if participants sent notes or messages to the study supervisors via the reporting feature of \system, these notes were also included for analysis. The analysis excluded chat messages between participants, as awareness of their messages being observed could introduce a bias in how they converse.
Reflexive thematic analysis similar to ~\citet{terry2017thematic} was used to analyze the data. Two authors inductively coded the interviews. This resulted in two codebooks (49 and 68 codes) that were merged by discussing similar codes and conflicting views (65 codes). In a joint session, open codes were then grouped into nine clusters, which resulted in three main themes. The themes are presented in the following, supported by excerpts from the interviews.\\

\subsubsection{Theme 1 - Integrating Remote Gardening while Navigating Daily Life}\label{subsubsec:TA-theme1} A core aspect of \system is its flexibility in allowing users to control their level of initiative. Participants first described their overall experiences and how they integrated \system into daily life. 

\subsubsection*{Routines}\label{subsubsec:TA-theme1-routines} They reported using \system at specific times in their routines, such as in the morning while brushing their teeth (\textit{P18}) or after work in the evening (\textit{P1-P4}). In contrast, irregular usage patterns were also reported (\textit{"I didn't check on it [their plot] at specific times, just randomly, sometimes at night even." P7}). When asked about the integration of \system into their routines and what facilitated it, participants highlighted easy access to a garden and being able to not only integrate \system into existing routines but also establish new routines around \system (\textit{"It wasn't much effort, so I just had a set time for it. At some point, it was almost like a routine to go check up [referring to their plot] straight after work." P1}).

\subsubsection*{Tailoring}\label{subsubsec:TA-theme1-tailoring} Participants gave various explanations for control mode switches. The novelty of the concept motivated participants to examine what FarmBot's capabilities (e.g., \textit{"I was curious. What does this thing [Farmbot] do? What will be the result?" P11}), but decisions to change control modes were also informed by schedules or breaks in them (\textit{"If I knew that my routine was going to be all over the place for some time, using fully automated would be the better choice for me since the plot gets taken care of no matter what" P1}). Apart from schedules, participants also considered personal aspects such as their gardening expertise when choosing their preferred control mode (e.g., \textit{"I don't know a thing [referring to gardening]. I kept it [\system] in fully automated at first. It would have taken way longer for me to think about when and what to do, or probably nothing would have grown" P5}). Rationales for the popularity of \textit{Hybrid} were grounded in perceiving it as a balanced trade-off (P2-P4, P6, P18). 

\subsubsection*{Disruptions}\label{subsubsec:TA-theme1-disruptions} Participants also shared how they leveraged \system's control modes to adapt their use to temporary changes to their daily routines. Such circumstances included general busyness (P10, P11), studying for an exam (P4), or traveling (P10, P11). For example, P4 switched to fully automated mode before an exam to free up more time for studying alongside work (\textit{"It's already quick [plant care], but I was somewhat overwhelmed before [exam name] and had to switch to more automation" P4}). External factors, such as weather conditions (P1, P2, P3, P15, P18), also shaped participants' ability to get involvement. During the second week and continuing into the third week of the study, frequent rain showers occurred, prompting a reduction in the frequency of gardening tasks (\textit{"And it rained at some point, so I stopped watering, and that created a bit of a break" P18}) and switching to \textit{Automated} so that checking if and when it stopped raining was not necessary (P2, P3, P6). 

\subsubsection{Theme 2 - Managing Distance with Digital Augmentation}\label{subsubsec:TA-theme2} \system was designed for fully remote operation, contrasting traditional in-situ gardening. Users must still manage gardening's inherent asynchronicity, where actions are connected to outcomes that unfold over time.

\subsubsection*{Asynchronicity}\label{subsubsec:TA-theme2-asynch} Participants felt that the monitoring and visualization features of \system were helpful as they provided an always-available way to check up on changes quickly (P1, P4, P6, P11, P15). Continuity in features like timelines was perceived as beneficial (\textit{"This photo concept with the timeline is pretty neat. I can go back and see what happened. That would actually be beneficial in real gardening but not just for progress. I could check for animals on my field too" P18}).
They additionally connected their preferred progress-tracking method to their chosen control mode. In \textit{Automated} mode, users mainly act as spectators, verifying that FarmBot performs tasks as expected to ensure successful crop cultivation. (\textit{"I concentrated primarily on the timeline. That's all I needed to know if the robot took care of my field" P5}). Conversely, utilizing more manual control modes prompted participants to focus more intently on observing plant growth to confirm the success of their actions (\textit{"I'm not very knowledgeable about this [gardening] and used it [photo grid] to try and judge if the plant is healthy, growing at the right pace, getting enough water, or too much sun." P12}).

\subsubsection*{Remoteness}\label{subsubsec:TA-theme2-remoteness} \system introduces a constant element of distance due to remote control. This was reported to impact the perception of actions and progress in the field, influencing the design of plots. Reasons for this were difficulties gauging the size of the field and how accurate measures were (P1-P4, P6, P12), leading to particular crop choices and layouts (\textit{"I know the app told me the dimensions, but I've not grown lettuce for example, so I didn't know how accurate it is and if it would all fit. I know that radishes don't get that big." P4}). Participants used features like live streams and map visualizations to keep track of their progress, yet plant growth is inherently slow. Until germination, the lack of visible changes, combined with the distance and inability to examine the field in situ, led to doubts and worry (\textit{"Well, once the seeds are in the ground, there isn't a lot going on. I saw the germination times, but still, at first, I thought, okay, did I do something wrong? Is it supposed to take this long?" P1}). 

\subsubsection*{Perception of Gardening}\label{subsubsec:TA-theme2-perception} General perceptions of how to care for a garden were also influenced, with participants sharing that they felt like they had to check their plot more often due to \textit{"having the garden in their pocket" (P18)}. Comparatively, gardening via \system was perceived to be distinctly different from traditional gardening, which participants described as more sensory-rich (\textit{"getting your hands dirty" (P10)}; \textit{"spending time in fresh air" (P11)}). Contrarily, \system was described as an alternative experience that emphasizes "\textit{providing an opportunity to do it [gardening] at all and get some healthy food." (P2)}. Participants found different ways to address the more distant gardening experience. For instance, manual control was highlighted as one such method as it allowed for more agency (\textit{"If I don't have time to be there, I’d say I still want to be in control, like, yes, I’m watering or weeding now, I’m the one pressing the button. I don’t need to decide everything myself, but I still want to have the final say, so it feels valuable for me." P12}). Participants also highlighted that they would have welcomed gamified elements such as streaks to further their motivation and attachment to the gardening process via \system (P7, P12, P15). 

\subsubsection{Theme 3 - Learning to Collaborate with a Robotic Gardening Partner in a Shared Space}\label{subsubsec:TA-theme3}
Despite participants having heard of more common smart home technology for gardening (e.g., mowing robots or automated irrigation), \system represented a novel experience.

\subsubsection*{Learning}\label{subsubsec:TA-theme3-learning} While \system aimed to lower the required gardening knowledge by delegating tasks to a PAR, it introduced new demands, such as technological proficiency (\textit{"Wrapping your head around it [collaborating with a robot] takes a while. I am not a tech-expert and use my phone to text at most." P3}). Using the \textit{Automated} mode allowed participants to observe the FarmBot in action and learn about its operations. By starting as spectators, they could gradually get involved in plant care. They described feeling more competent about plant care (P1, P4, P5, P7), which seemed to raise motivation and confidence (e.g., \textit{"If you've never really done it [gardening] before, you probably think you need a green thumb or some kind of special skills. But actually, I found that since I was able to watch and learn from the robot at the beginning and then do it independently in the end, I feel like I can do it." P4}). It also became apparent that UI issues such as delays become a much more potent risk in remote gardening. For instance, P1 described the following: \textit{"I think I overwatered my field at the beginning. I pressed the "Water All" button, but it took a bit to give me feedback, so I pressed it 2-3 more times. I then saw that all my tapping was taken as individual watering requests."}. While this has the potential to cause destruction, it has also led to more extensive reflection before deciding on what action to execute (P1, P4, P12, P15, P18).

\subsubsection*{Comparison}\label{subsubsec:TA-theme3-comparison} Sharing one PAR with other gardeners also influenced participants. While some participants did not pay extensive attention to the other plots around them, others were influenced more profoundly. \system implements no competitive gamification elements (e.g., leaderboards). Yet, participants who did monitor growth on the entire field (e.g., using the map or live streams) described feelings of competitiveness from comparing plot designs and growth progress. This prompted some users to plant more than they had originally envisioned (\textit{"I planted some seeds at the beginning but then again around week two. The plot next to me [P8] had way more and green bits were already showing" P7}). In contrast, observing growth across the entire field led to satisfaction and joy toward the conclusion of the study (\textit{"The most positive thing for me was that at some point you could see when everything started to grow. Took a while but then you feel like there's quite a lot coming suddenly." P15}).

\subsubsection*{Dissent} \label{subsubsec:TA-theme3-dissent}
Participants generally trusted FarmBot to perform tasks as requested, but concerns arose over factors like water usage and watering height (P1, P4, P6, P15, P18), often reflecting how they would act in the garden themselves. (\textit{"The watering looked pretty aggressive to me. I don't know if that's good for those smaller seedlings. Personally, I would have been a bit more careful with that." P15}). While participants rarely interacted with \system simultaneously, situations where FarmBot had to handle multiple garden care requests did arise. Participants described situations where this led to longer waiting times, as FarmBot was handling tasks of all fully automated users (cf. \autoref{subsubsec:control-strategies}), while those using \textit{Manual} or \textit{Hybrid} mode logged in to execute their tasks (e.g., \textit{"I wanted to water my plants in the evening because I thought it would be better for them, but I saw that the robot's task queue had gotten quite long. I guess that's when the robot took care of the plants of all users. [..] I submitted my request and left. Would have taken too long to wait and watch." P12}).

\subsubsection{Positive \& Negative Aspects}\label{subsubsec:open-comments}
In the open feedback, participants mentioned several positive and negative aspects of \system. The system’s flexibility in switching between various modes, enabling gradual learning and control over the gardening process, was appreciated (1x). \system was described as easy and intuitive to use (1x), simple, and time-saving (3x), with participants enjoying the connection between robotics and gardening, comparing it to \textit{"Tamagotchi for grown-ups" (P8)}.  Participants also highlighted the live stream feature and remote watering functionality, which increased their sense of control and involvement (1x). Interaction with \system further sparked new motivation for participants to improve their gardening expertise by cultivating indoor plants (4x).
Areas for improvement included a preference for using \system on laptops or PCs for better visibility (3x) and the lack of notifications in automated mode, leading to disengagement (5x). More feedback was requested (2x), such as interface cues (e.g., inactive buttons) and detailed task performance metrics (e.g., water usage in milliliters).

\section{Discussion}
In the following, we discuss our study outcomes and reflect on the design process of \system. Based on our insights, we propose design considerations for remote robot-assisted urban gardening experiences.
\subsection{Design \& Field Deployment of \system}
\subsubsection{Design Goals}
To address practical barriers preventing interested citizens from pursuing urban gardening, \system's design was guided by three goals (cf. \autoref{subsubsec:design-goals}): enabling on-demand multi-user interaction, supporting dynamically adjustable levels of involvement, and providing experiences that extend beyond basic plant care. To promote accessibility (\textit{D1}), \system was implemented as a smartphone-optimized web application, allowing users to perform tasks and monitor the field in real time. Login data coupled with qualitative feedback, indicated that \system facilitated these activities, ensuring accessibility at any time. In alignment with \textit{D2}, \system offered three control modes (cf. \autoref{subsubsec:control-strategies}), enabling users to adjust their level of involvement based on their preferences and schedules. User logs revealed transitions between modes, reflecting flexibility and adaptability, while interview responses demonstrated integration into daily routines.
\system further incorporated features like time-lapses, growth visualizations, a timeline, and chat functionality to foster interactions beyond plant care execution (\textit{D3}). User logs showed engagement with most features, excluding the chat, throughout the study. Study results additionally showed a significant increase in perceived connectedness and personal conservation perception.

\subsubsection{Gardening Outcomes}
Previous research has shown that food production is among the most common motivators for citizens to engage in urban gardening~\cite{goodfellowbarriers2022, opitz2016}. In our evaluation of \system, we explored how effectively users could cultivate their desired crops within their garden spaces despite the unfamiliar notion of collaborating with a PAR. Following calls in prior research~\cite{Webber2023CHI}, we deployed \system under real circumstances (i.e., freestanding). Overall, the success rate and gardening outcomes were satisfactory (cf. \autoref{subsec:results-garden-cultivation}). Sixteen participants continued using \system beyond the study period to maintain their garden. From qualitative responses in the interview, we conclude that participants did not see \system as a replacement for traditional gardening, as it lacks the sensory-rich experience of direct contact with a garden. Instead, they viewed it as a convenient alternative that makes gardening more accessible (cf. \autoref{subsubsec:TA-theme2-perception}). 

\subsubsection{Connectedness}
Discussion on whether technology enhances or diminishes nature-based activities like gardening has been a long-standing topic in HCI literature~\cite{Rodgers2020, jones2018dealing}. With systems like \system, which enable fully remote interaction and allow users to offload gardening responsibilities to a PAR, acting as a \textit{permanent garden collaborator}, there is a potential trade-off between accessibility, convenience, and the authenticity of nature experiences. Building on prior research (e.g.,~\cite{reflection1, mediatedNature, wesenseEdu}) and inspired by the FarmBot web application, we implemented strategies that allowed users to monitor garden conditions, reflect on past developments via time-lapses and timelines, and anticipate future growth using visual indicators. Pre- and post-study IOS ratings suggest that \system enhanced participants' connection to their plots, possibly due to first-time gardening experiences or renewed interest. Trends by preferred PAR control method suggest that participants using the \textit{Automated} mode reported lower connectedness, while those favoring \textit{Hybrid} or \textit{Manual} modes reported higher connectedness. Facilitation of routine-building was a common qualitative finding and has been connected to lasting engagement in prior literature~\cite{prochaska2001stages}. We argue that offering this spectrum of control is likely beneficial for engaging people in urban gardening. However, it also raises the need for further exploration in HCI to understand how interface design can be refined to prevent disengagement caused by over-reliance on convenience.

\subsubsection{Remote Gardening as an Entry Point} A common barrier to engagement in urban gardening is negative past experiences~\cite{lewis2018, conwaybarriers2016c} or limited gardening expertise, making gardening seem difficult or unapproachable. Prior approaches already address this point by providing information~\cite{growkit,wesenseEdu} and using automation~\cite{pearceSmartWatering}. \system extends this by allowing users to use automation based on their personal desire for involvement. We found that novice participants used this concept to gradually gain confidence by watching FarmBot perform tasks. Contrarily, experienced gardeners sometimes questioned how FarmBot was executing tasks (e.g., watering) and grounded their skepticism in their expertise. Aligning with existent research (e.g.,~\cite{lewis2018role}), for more experienced gardeners, designing with an emphasis on trust gains more relevance. This can include allowing more fine-grained control over PAR parameters, such as watering height.  However, it should be noted that while \system reduces demands for required knowledge to garden, it introduces new requirements, such as technical proficiency with a smartphone and a willingness to gain a basic understanding of how a PAR operates. Existent literature highlights that this may, in itself, represent a new barrier~\cite{proficiencybarrier1} that may diminish inclusivity, increasing the importance of intuitiveness for fully remote gardening concepts similar to \system. 

\subsubsection{Intergardener Effects}
\system was designed to leverage one PAR shared among multiple users. Additionally, the provided global map allows gardeners to review the entire field (i.e., other gardeners' progress). Based on interview responses, this led to varying outcomes. For one, gardeners shared joy for each other's progress, finding it pleasant and satisfying to follow. However, the ability to review and judge other gardeners' progress also invited participants to compare their plot layout and progress to that of other participants. This introduced aspects of competitiveness, replicating known group behaviors~\cite{graziano1997competitiveness}. Unlike community gardens, where all actions can be viewed by present members, virtual gardens that are only worked remotely provide the opportunity to hide certain information, such as progress or executed actions. These aspects are currently not addressed by \system, as all participants knew they could review the progress on the entire field. Hiding information gardeners feel uncomfortable sharing could further reduce feelings of insufficiency and pressure to achieve a positive gardening outcome. 

\subsubsection{Engagement Over Time} Deploying \system for three weeks enabled us to observe user behavior beyond the initial novelty phase. After bulk seeding and exploring \system's features, participants adapted to regular usage. While login data indicates frequent and consistent access, analysis of control mode switches and UI interactions reveals a shift from active engagement to passive monitoring over time.
\system connects virtual actions with tangible changes in a distant garden, creating a bi-directional relationship where participants influence the physical environment while also being affected by external factors (e.g., rainfall reducing the necessity of plant care actions). Such disruptions sometimes led to disengagement. Gamification (e.g., streaks) and notifications were noted to achieve re-engagement. We intentionally avoided this approach, as our target audience consisted of users already interested in gardening. However, based on these results, we conclude that incorporating digital strategies to motivate engagement is advisable to address the perceived distance of systems like \system.

\subsection{Design Considerations \& Implications}
The following paragraphs present design considerations we established based on the development process of \system and the insights from our 3-week deployment.

\subsubsection{Encourage and Support Free Exploration}
In in-situ gardening experiences, exploration is an element that is ever-present and can support learning and encounters with other gardeners~\cite{wesenseEdu} or non-human actors~\cite{CameraTrap}. Behaviors, where users can freely explore their and other gardeners' spaces, are equally as important in remote gardening experiences. Prior work proposing distant nature experiences has typically enabled exploration of environments at different representation levels (i.e., \textit{abstracted, mediated, simulated})~\cite{Webber2023CHI}. \system embraces this notion by integrating views that show the real environment, coupled with more abstracted or generated visualizations. The offered views were, however, predefined and static in that cameras could not be moved except for those attached to FarmBot. This allows users to explore only a fraction of the actual space. Open comment suggestions indicate that using 360° video players that allow for more extensive exploration would be preferred over static views. This could further users' interest and provide improved situation awareness~\cite{litReviewHRI}.

\subsubsection{Embrace Risk as a Design Element to Foster Connectedness}
When novices start gardening, it is common for initial attempts to fail due to mistakes. While this can lead to demotivation~\cite{lewis2018,conwaybarriers2016c}, it is also an inherent part of the learning process. Given that \system connects virtual actions to the tangible shaping of a garden space, there is a constant risk that destruction may occur. For instance, when users believed they had made a mistake, such as repeatedly watering, it caused initial worry but also increased carefulness in following interactions. The risk of irreversible destruction as a design element has been proposed as a design element in prior research (e.g.,~\cite{dungeonmaker, beatrossmy}). To foster attachment and care, we argue that deliberately balancing this risk factor, allowing for some mistakes (as \system does in its \textit{Manual} mode), should be considered more broadly in PAR-supported remote gardening and distant nature engagement concepts.

\subsubsection{Consider Sustainability when Resource-Intensive Actions Become Easily Accessible}
Sustainable gardening practices emphasize conserving natural resources, promoting biodiversity, and minimizing environmental impact. In our deployment of \system, we observed that distance when interacting with a garden remotely can lead users to perceive it as more game-likely, as reflected in open feedback comparing it to \textit{"Tamagotchi for grown-ups" (P8)}. Coupled with the notion of having "\textit{the garden in your pocket" (P18)}, such views can lead to phases where users experiment with different interface elements, such as repeatedly moving the robot or excessively watering before planting seeds, simply to observe the system’s responses. While this experimentation can boost user engagement, the waste of resources such as water in a distant location may not be as present. This is especially true when the system allows for more experimental interactions, where users have full control, and the robot takes a secondary role. We propose introducing resource management systems to minimize resource consumption, such as limiting daily water use or robot movement requests. While users may never use resources to their limit, the visual indication of limitations can serve as a nudging mechanism similar to eco-feedback~\cite{ecofeedback1} to avoid waste and foster a sense of conservation for the collective, which reflects suggestions from sustainable HCI work (e.g.,~\cite{mencarini}). 

\subsubsection{Leverage Digital Augmentation to Reveal Hidden Aspects of Plant Growth}
Gardening is inherently a slow process that requires patience, often involving phases where no visible changes occur in the garden bed. In shared or community gardens, social interactions often bridge these idle phases~\cite{community}. With \system, user logs revealed that during the early stages, when users typically planted their seeds, there was an increase in visualization changes, information-seeking behaviors, and feelings of \textit{"wanting something to happen"}. Since no immediate physical changes were visible in the garden, participants sought alternative ways to stay informed about the garden’s status. Typically, underground growth processes are hidden, but the sensing capabilities of PARs, coupled with real-time visualizations, could bridge this gap by digitally augmenting and displaying information about usually hidden growth processes. 
\section{Limitations \& Future Work}
\subsubsection*{Limitations}
We acknowledge several limitations of our study. Our field deployment of \system was designed to provide participants with a realistic scenario in which \system could be used in the future. To achieve this, we installed the farming robot controlled by \system in a real garden bed and deployed it for three weeks. However, this timeframe does not fully capture the gardening process, which typically spans entire seasons with different crops sown at various times. While we allowed participants to retain access to their garden plots beyond the study, with 16 participants opting to do so, the data from this extended period was not included in our analysis. As a result, ongoing engagement with \system and the sustained positive effects may have evolved differently if the study had been longer and our findings should be viewed early findings. Another limitation lies in the farming robot itself. \system envisions a future where such robots are widely available in urban environments, potentially serving as caretakers of green spaces~\cite{goddard2021}. However, this concept is still speculative, and the current deployment relies on a single robot in a supervised setting. This limits the generalizability, as scaling the system to larger urban contexts with multiple robots could introduce new challenges in terms of coordination, maintenance, and user interaction. Lastly, four participants in our study sample had access to a green space where they could grow their own crops and utilized the \system field in a manner similar to an allotment garden. We acknowledge that these participants do not fully represent urban residents who are unable to engage in gardening due to a lack of suitable spaces, thereby reflecting an alternative use case. 

\subsubsection*{Future Work}
We optimized PlantPal for mobile phones to ensure proper content display across various screen sizes. However, issues arose with older smartphone models featuring smaller screens, where content was not rendered correctly. Future iterations of \system should address these limitations, including optimization for larger screens, as requested by participants in our study. In our study, novice gardeners often observed the robot to learn gardening processes, which increased their confidence. This approach could be expanded to other contexts, such as traditional community gardens. Similar to research using drones to teach movement patterns~\cite{drones4dance}, PARs could support novice gardeners as they build foundational skills in these settings. Additionally, our evaluations revealed trends indicating that the level of control significantly influences the gardening experience. Conducting sufficiently powered studies to compare different control modes and their longitudinal impact on the gardening experience can deliver additional implications for design and contexts in which each is suited best. 

\section{Conclusion}
In this work, we proposed using remote collaboration PARs to enable garden cultivation regardless of location, space limitations, or time constraints. By reviewing the literature on HRI, urban gardening, and HCI, we identified the capabilities of current PARs and how they align with the challenges citizens face in urban gardening. We conducted a formative survey to sample preferences regarding our concept and cleared up design ambiguities Building on these insights, we developed \system, a web application that provides on-demand access to a garden space via a PAR. \system tackles urban gardening barriers by enabling flexible user involvement and enhancing engagement with digital augmentations. To evaluate \system, we deployed it on a real garden bed (18m²) for three weeks with N=18 participants. Participants were able to cultivate their own garden space successfully. They reported having had an enjoyable experience where they could establish a connection with their plot despite the remote setting. We derived design considerations addressing exploration, risk, sustainability, and digital augmentation based on our findings. These considerations can broadly inform the design of future PAR-enabled urban gardening concepts and fully remote technology-supported nature experiences.

\section*{Open Science}
We make the source code for the \system web application, 3D models and scripts for our extended camera system, and hardware-related insights/manuals publicly available. They can be accessed at https://github.com/J-Britten/PlantPal.

\begin{acks}
We thank the anonymous reviewers for their constructive and insightful feedback. We also extend our gratitude to Prof. Dr. Steven Jansen, Peter Zindl, Markus Wespel, and the staff of the Botanical Garden Ulm for their collaboration and support throughout this project. This work was funded by the Landesgraduiertenförderung (LGFG) Scholarship for Ph.D. students.
\end{acks}
\bibliographystyle{ACM-Reference-Format}
\bibliography{references}

\appendix

\section{Guideline Protocol Semi-Structured Interviews}\label{appendix:A}
\subsection*{1. General Usage}
\begin{itemize}
    \item Describe your general experiences from week 1 to week 3.
    \begin{itemize}
        \item What were your goals?
        \item What was the focus over the weeks?
    \end{itemize}
    \item What was the most enjoyable aspect? What was the least enjoyable?
    \item Describe a typical day and how the app was integrated into it.
    \item What was your thought process behind creating the garden layout?
    \item To what extent did the fact that it was a virtual field influence your decisions?
    \item During the study, did you observe the state of other participants' fields? Why or why not?
\end{itemize}

\subsection*{2. Levels of Automation}
\begin{itemize}
    \item How were the different levels of automation perceived?
    \item Did you switch between levels of automation?
    \begin{itemize}
        \item Why or why not?
        \item What triggered a switch?
    \end{itemize}
    \item Which level of automation was the most useful? Why?
    \item Which level was the least useful? Why?
\end{itemize}

\subsection*{3. Effects of Remote Interaction}
\begin{itemize}
    \item How did it feel to never have to be on-site?
    \item To what extent did the app’s features compensate for not being physically present?
    \item How was the support for decision-making (e.g., what to plant, when to water/whether watering was needed) perceived?
    \item What was observed in the live streams/images?
    \begin{itemize}
        \item What was the goal of the observation?
        \item What specifically did you see?
        \item Were there any surprises?
    \end{itemize}
\end{itemize}
\subsection*{4. Interaction with the Robot}
\begin{itemize}
    \item How did you assess the robot’s capabilities?
    \item How was it perceived that the robot was a shared resource for all fields?
    \item Did you observe the robot while it was performing its tasks? Why or why not?
    \item Were you afraid of breaking something? Why?
    \item Were there any problems, confusion, or surprises regarding the robot's behavior?
    \item To what extent did participating in the study and using the system influence your perception of your gardening skills?
\end{itemize}


\end{document}